\newcommand{\teff}{T$_{\mathrm{eff}}$}
\newcommand{\Teff}{T$_{\mathrm{eff}}$}
\newcommand{\logg}{log $g$}

\newcommand{\vmicro}{$\rm{v_{micro}}$}

\newcommand{\project}[1]{\textsl{#1}}
\newcommand{\gaia}{\project{Gaia}}

\documentclass[a4paper,fuseAMS,leqn,usenatbib]{mnras}

\usepackage{graphicx}
\usepackage{amssymb}
\usepackage{amsmath}
\usepackage{times}
\usepackage{subfigure}
\usepackage[T1]{fontenc}
\usepackage{ae,aecompl}
\usepackage{textcomp}

\usepackage{hyperref}

\title[COMBS II]{The COMBS Survey - II. Distinguishing the Metal-Poor Bulge from the Halo Interlopers\thanks{Based on observations collected at the European Southern Observatory under ESO programme: 089.B-069 }}

 \author[Lucey, et al. 2020]{Madeline~Lucey$^{1}$\thanks{E-mail:m\_lucey@utexas.edu}, Keith~Hawkins$^{1}$, Melissa~Ness$^{2,3}$, Victor~P.~Debattista$^4$, \newauthor Alice~Luna$^1$,
 Martin~Asplund$^{5}$,  
 Thomas~Bensby$^{6}$, 
 Luca~Casagrande$^{7}$, 
\newauthor Sofia~Feltzing$^{6}$, 
 Kenneth~C.~Freeman$^{7}$,  
 Chiaki~Kobayashi$^{8}$ 
 and Anna~F.~Marino$^{9,10,11}$
\\
$^{1}$Department of Astronomy, The University of Texas at Austin, 2515 Speedway Boulevard, Austin, TX 78712, USA \\
$^{2}$Center for Computational Astrophysics, Flatiron Institute,
162 5th Ave., New York, NY 10010, USA\\
$^{3}$Department of Astronomy, Columbia University, 550 W 120th St., New York, NY, 10027, USA \\
$^4$Jeremiah Horrocks Institute, University of Central Lancashire, Preston, PR1 2HE, UK \\
$^5$Max Planck Institute for Astrophysics,  Karl-Schwarzschild-Str. 1, D-85748 Garching, Germany \\
$^{6}$Lund Observatory, Department of Astronomy and Theoretical Physics, Box 43, SE-221\,00 Lund, Sweden \\
$^{7}$Research School of Astronomy and Astrophysics, The Australian National University, Canberra, ACT 2611, Australia \\
$^{8}$Centre for Astrophysics Research, Department of Physics, Astronomy and Mathematics, University of Hertfordshire, Hatfield AL10 9AB, UK \\
$^9$Dipartimento di Fisica e Astronomia “Galileo Galilei”, Universita' di Padova, Vicolo dell’Osservatorio 3, I-35122, Padua, Italy \\
$^{10}$Istituto Nazionale di Astrofisica - Osservatorio Astronomico di Padova, Vicolo dell’Osservatorio 5, IT-35122, Padua, Italy \\
$^{11}$Centro di Ateneo di Studi e Attivita' Spaziali “Giuseppe Colombo” - CISAS, via Venezia 15, I-35131, Padua, Italy}

\date{Accepted . Received ; in original form }

\pubyear{2020}

\begin{document}
\label{firstpage}
\pagerange{\pageref{firstpage}--\pageref{lastpage}}
\maketitle

\begin{abstract}
The metal-poor stars in the bulge are important relics of the Milky Way's formation history, as simulations predict that they are some of the oldest stars in the Galaxy. In order to determine if they are truly ancient stars, we must understand their origins.  Currently, it is unclear if the metal-poor stars in the bulge ([Fe/H] < -1 dex) are merely halo interlopers, a unique accreted population, part of the boxy/peanut-shaped bulge or a classical bulge population. In this work, we use spectra from the VLT/FLAMES spectrograph to obtain metallicity estimates using the Ca-II triplet of 473 bulge stars (187 of which have [Fe/H]<-1 dex), targeted using SkyMapper photometry. We also use \gaia\ DR2 data to infer the Galactic positions and velocities along with orbital properties for 523 stars. We employ a probabilistic orbit analysis and find that about half of our sample has a > 50\% probability of being bound to the bulge, and half are halo interlopers. We also see that the occurrence rate of halo interlopers increases steadily with decreasing metallicity across the full range of our sample (-3 < [Fe/H] < 0.5). Our examination of the kinematics of the confined compared to the unbound stars indicates the metal-poor bulge comprises at least two populations; those confined to the boxy/peanut bulge and halo stars passing through the inner galaxy. We conclude that an orbital analysis approach, as we have employed, is important to understand the composite nature of the metal-poor stars in the inner region.

\end{abstract}


\section{Introduction}

\label{sec:Introduction}

Piecing together the history of our Galaxy, the Milky Way (MW), is one of the major objectives of astrophysics and will lead to new insights in our understanding of galaxy evolution in general. The center of our Galaxy is one of the least understood components given that it has historically been difficult to study. High levels of both crowding, which makes it difficult to resolve individual stars, and of extinction, which makes it difficult to achieve high signal-to-noise ratio data have prevented substantial studies of the Galactic bulge until recently. 

Large spectroscopic surveys such as Bulge Radial Velocity Assay \citep[BRAVA,][]{Rich2007}, the Abundances and Radial velocity Galactic Origins Survey \citep[ARGOS,][]{Freeman2013},
the GIRAFFE Inner Bulge Survey \citep[GIBS,][]{Zoccali2014}, the HERMES Bulge Survey \citep[HERBS,][]{Duong2019}, the Extremely Metal-poor BuLge stars with AAOmega survey \citep[EMBLA,][]{Howes2015}
and the Apache Point Observatory Galactic Evolution Experiment \citep[APOGEE,][]{GarciaPerez2019,Rojas-Arriagada2020}  have measured the radial velocities and chemical abundances of bulge stars. One of the major results from these surveys is the measurement of the metallicity distribution function (MDF) of the central part of the MW. The ARGOS survey, which used 14,150 stars, determined that the MDF is made up of five components \citep{Ness2013a}. They associate the five components with different components of the Galaxy. The highest metallicity components (peaks at [Fe/H]= +0.15 and -0.25 dex) they associate with the boxy/peanut-shaped (B/P) bulge. The three most metal-poor components they associate with the thick disk (peak at [Fe/H]=-0.7 dex), the metal-weak thick disk (peak at [Fe/H]=-1.18 dex) and the stellar halo (peak at [Fe/H]=-1.7 dex). However, the higher metallicity components dominate with only 5\% of stars having metallicities < -1 dex \citep{Ness2016}. Other studies have found similar results with slight variations \citep[e.g.,][]{Zoccali2008,Johnson2013a,Zoccali2017,Bensby2013,Rojas-Arriagada2014,Bensby2017,Rojas-Arriagada2017,Duong2019}. However, \citet{Johnson2020} find that the multi-peak model is only valid for the outer bulge and within a Galactic latitude ($b$) $\sim6^{\circ}$ the metallicity distribution is best described by a single peak with a long metal-poor tail, consistent with a closed box model.

Although they only comprise a small fraction of the bulge, the metal-poor stars have become of particular interest recently. Simulations have shown that the metal-poor stars in the center of the Galaxy may hold critical information about the first stars and early Galaxy evolution. For example, simulations predict that if Population III stars exist in our Galaxy, they 
are more likely to be found in the bulge \citep{White2000,Brook2007,Diemand2008}. It has also been predicted that stars of a given metallicity are typically older if they are found in the center of the Galaxy  \citep{Salvadori2010,Tumlinson2010,Kobayashi2011a}. Furthermore, simulations show that if one selects metal-poor stars, then the fraction of the oldest stars becomes highest towards the Galactic center \citep{Starkenburg2017a, El-Badry2018b}. Therefore, targeting metal-poor stars towards the center of the Galaxy is conducive for the discovery of ancient stars.

However, discovering metal-poor stars that are currently in the bulge is not enough to assume they are ancient. These stars have many possible origins which may correspond to different age distributions. For example, it is unclear if these stars are confined metal-poor bulge stars that stay confined to the bulge or if they are halo stars that are just passing through the bulge and actually spend most of their time at large distances from the Galactic center. If they do stay confined to the bulge, it is uncertain if they are a classical bulge population or participate in the B/P bulge. The signature of a classical bulge is a pressure-supported component that is the result accretion in the hierarchical growth of galaxies model \citep{Kauffmann1993,Guedes2013} or is the rapid assembly of gas-rich small sub-galaxies \citep{Kobayashi2011a}. On the other hand, a B/P  bulge is rotation-supported and formed through secular evolution of the bar either by buckling instabilities \citep{Raha1991,Merritt1994,Bureau2005,Debattista2006} or orbit trapping \citep{Combes1981,Combes1990,Quillen2002,Quillen2014,Sellwood2020}. Most of the mass in the bulge has been shown to participate in the B/P bulge \citep{Howard2009,Shen2010,Ness2013b,Debattista2017}. However, it has been suggested that the MW has a compound bulge \citep[a B/P bulge with a classical bulge;][]{Athanassoula2005} where the less massive metal-poor component is a classical bulge population \citep{Babusiaux2010,Hill2011,Zoccali2014}.  As a B/P bulge and a classical bulge are the result of different formation histories, it is essential to distinguish between these scenarios in order to determine if these stars are truly ancient. On the other hand, if these stars do not stay confined to the bulge, then it is possible that they are part of a unique accreted population or the in-situ halo. Consequently, it is essential to study the chemistry and kinematic properties of the metal-poor stars in the bulge in order to distinguish between these possible origin scenarios and determine whether they are truly the oldest stars in the Galaxy. 

To this end, there have been a number of studies on the chemistry of metal-poor bulge stars. The first installment of the Chemical Origins of Metal-poor Bulge Stars (COMBS) survey studied the detailed chemical abundances of 26 metal-poor bulge stars \citep{Lucey2019}. One of the main results from this work is that the metal-poor bulge has higher levels of Calcium (Ca) enhancement compared to the disk and halo. In addition, the metal-poor stars have lower dispersion in the $\alpha$-element abundances (Ca, Silicon, Magnesium and Oxygen) than halo stars of similar metallicity. These results indicate that either metal-poor bulge stars are not halo stars and are a unique Galactic population or that the halo is more chemically homogeneous towards the Galactic center. The HERBS survey found complementary results \citep{Duong2019}. They also observed higher levels of Ca-enhancement and lower dispersion in the $\alpha$-elements for metal-poor bulge stars. In addition, the Carbon and neutron-capture material abundances have shown deviations from the halo distributions. Carbon-Enhanced Metal-Poor (CEMP) stars occur at a rate of 15-20\% among halo stars with [Fe/H]<-2 dex \citep{Yong2013}. However, we know of only one that has been observed in the bulge \citep{Koch2016}. 
After accounting for mixing that occurs during the red giant branch phase, the EMBLA survey found one out of 23 stars with [Fe/H] <-2 dex may have had a natal [C/Fe] >1 dex \citep{Howes2015}. Although, the lack of CEMP stars in the EMBLA survey could at least partially be a selection effect from the SkyMapper photometry \citep{DaCosta2019}. Similarly, neutron-capture enhanced stars have been observed at a lower rate than in the halo  \citep{Johnson2012,Koch2019,Lucey2019,Duong2019}.

Studies of the kinematics of metal-poor bulge stars indicate that they are distinct from the metal-rich population and do not participate in the B/P bulge. Using the line-of-sight velocities, it has been shown that the metal-poor component of the bulge rotates slower than the metal-rich component and has higher velocity dispersion \citep{Ness2013b,Kunder2016,Arentsen2020}. Furthermore, the vertex deviation, which measures the orientation of the covariance between the radial and tangential motion, approaches zero for metal-poor bulge stars while it is large for metal-rich stars \citep{Soto2007,Babusiaux2010}. This indicates that the metal-poor stars do not participate in the bar structure since the vertex deviation is large for a triaxial bar and zero for a stationary axisymmetric disk \citep{Zhao1994}. These observations are typically interpreted as evidence for a classical bulge population. However, \citet{Debattista2019} demonstrated that a vertex deviation of zero for metal-poor stars does not necessarily indicate an ex-situ classical bulge population. Furthermore, it is important to be careful when interpreting these previous results on the metal-poor bulge because it is unclear how many of these stars are confined bulge stars or are merely halo interlopers. For example, \citet{Howes2015} found that roughly half of their very metal-poor bulge stars ([Fe/H]$<-2$) had orbits bound to the bulge. Using RR Lyrae stars, \citet{Kunder2020} separated the halo interlopers from the confined stars and found evidence for a B/P bulge and a classical bulge population.

In this work, we aim to remove the interlopers from our sample using orbit analysis in order to determine the properties of confined metal-poor bulge stars. We present metallicity estimates for 473 stars (187 of which have [Fe/H] <-1 dex) and 3D kinematics for 523 stars, all of which are stars near the Galactic bulge.  In Section \ref{sec:data} we present the data we use to accomplish this work. We describe the method for determining the metallicities from the Ca-II triplet (CaT) in Section \ref{sec:metallicities}. The derivation of the kinematics and orbital properties is outlined in Section \ref{sec:dynamics}. Last, we discuss the fraction of metal-poor stars in the bulge that stay confined to the bulge in Section \ref{sec:confined} and the properties of the stars that do stay confined in Section \ref{sec:properties}.

\section{Data} \label{sec:data}

\begin{figure}
    \centering
    \includegraphics[width=\columnwidth]{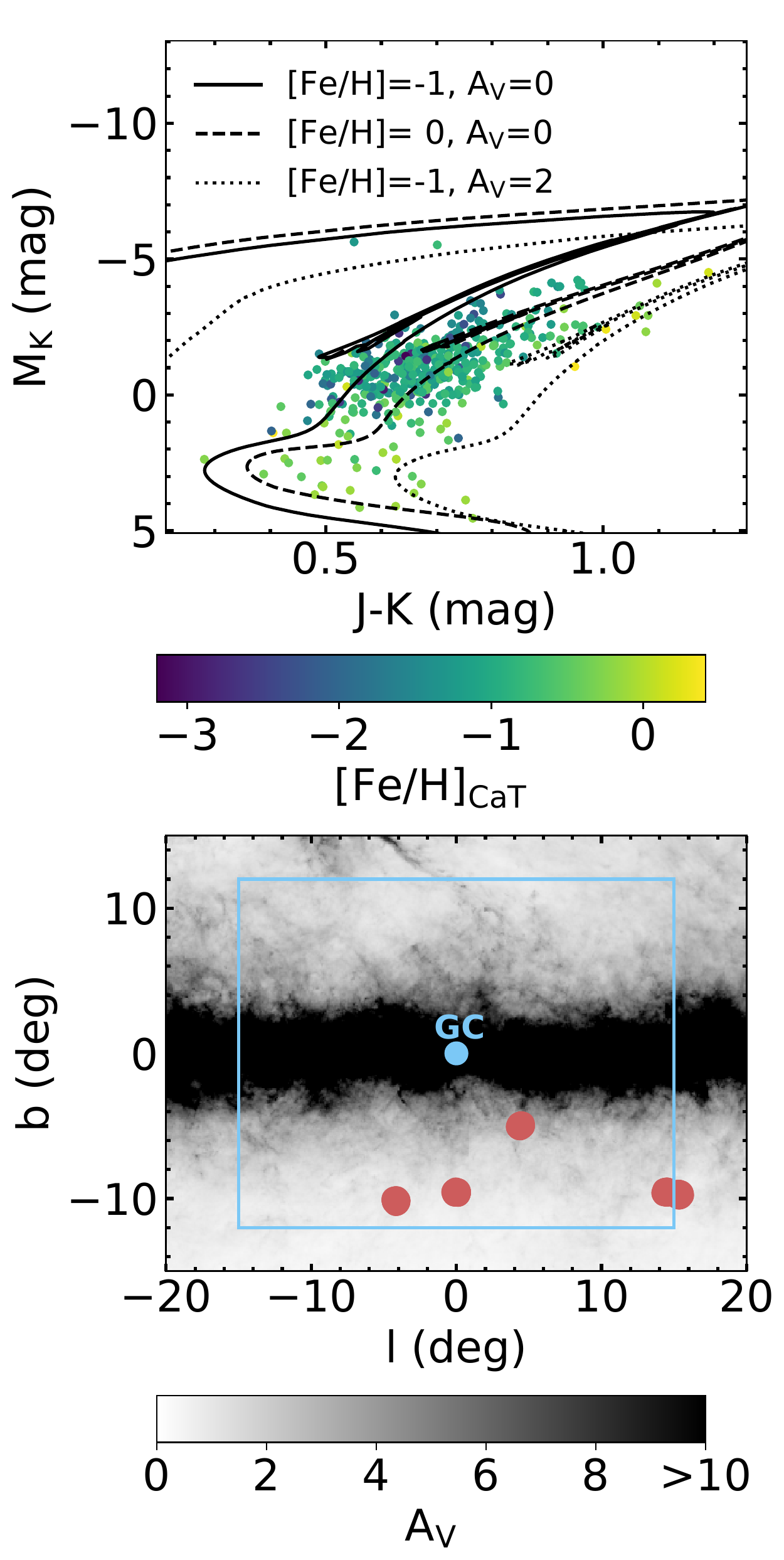}
    \caption{In the top plot, we show a color magnitude diagram of our sample colored by metallicity. On the y-axis we show the absolute K-band magnitude which is determined using our derived distance estimates. We only use ``A" quality photometry from the 2MASS survey \citep{Skrutskie2006}. For comparison, three isochrones with age 10 Gyr and varying metallicities and extinctions are shown in black. In the bottom plot, we show the Galactic longitudes and latitudes for the fields in our survey as red points. We also show the extinction map from \citet{Planck2013dust} in the background. A box roughly indicating the bulge region and a point indicating the Galactic center (GC) are shown in blue. }
    \label{fig:data}
\end{figure} 

Historically, observing large numbers of metal-poor bulge stars has been difficult given that they only make up around 5\% of stars in the Galactic bulge \citep{Ness2016}. However, photometric surveys, like the SkyMapper survey, which has a filter set designed to provide accurate stellar parameters \citep{Keller2007,Casagrande2019}, enabling the detection of extremely metal-poor stars for spectroscopic follow-up \citep[e.g.,][]{Keller2014,Howes2015}. Our stars have been selected using SkyMapper photometry along with ARGOS spectra \citep{Freeman2013} to target metal-poor stars within 3.5 kpc of the Galactic center. Our selection was made using uncalibrated commissioning photometry which is not included in the SkyMapper data releases as the pipeline is not yet optimized to deal with high levels of crowding. However, the use of this photometry suffices for the selection of metal-poor stars. For more details about the selection method, we refer the reader to Section 2 of \citet{Lucey2019}. 

Our spectroscopic data were obtained using the FLAMES instrument \citep{Pasquini2002} on the European Southern Observatory's (ESO) Very Large Telescope (VLT). We use the MEDUSA fibers, which feed to the GIRAFFE spectrograph along with the UVES spectrograph fibers. Therefore, we have high resolution data (R=$\lambda/\Delta \lambda \sim$ 47,000) from the UVES spectrograph along with medium resolution data (R$\sim$ 20,000) from the GIRAFFE spectrograph. We observed 40 stars with the UVES spectrograph and 555 stars with the GIRAFFE spectrograph, prioritizing the most promising metal-poor targets for the high resolution data. 

In top plot of Figure \ref{fig:data}, we show the color magnitude diagram of our sample. We only use ``A" quality photometry from the 2MASS survey \citep{Skrutskie2006}. We color the points by the metallicity that we derive in Section \ref{sec:metallicities}. The distances we derive in Section \ref{sec:dynamics} are used to convert the apparent K-band magnitudes into absolute magnitudes. For comparison, we also show MIST isochrones with age 10 Gyr \citep{Paxton2011,Paxton2013,Paxton2015,Dotter2016,Choi2016}. Specifically, we show an isochrone with [Fe/H]=-1 dex, $\rm{A_V}$=0 mag (black solid line), [Fe/H]=0 dex, $\rm{A_V}$=0 mag (black dashed line) and [Fe/H]=-1 dex, $\rm{A_V}$=2 mag (black dotted line). The majority of our stars have magnitudes consistent with red giant stars, red clump stars, or horizontal branch stars. The spread in color is due to a combination of varying metallicities and levels of extinction. The more metal-rich stars are generally redder than the more metal-poor stars. However, we do not de-redden the photometry. Therefore, the varying levels of extinction causes the metal-poor stars to appear redder and obscure the relation between color and metallicity. We also have a number of stars whose magnitudes are consistent with sub-giant stars. These stars are generally more metal-rich and are likely contamination from the disk along the line of sight towards the bulge. There are two stars whose magnitudes are consistent with planetary nebula. However, it is likely that these bright absolute magnitudes are the result of overestimated distances. Both of these stars have negative parallaxes and estimated distances > 20 kpc. However, these stars also have large distance errors, with the low error bar putting them within a distance of 11-14 kpc. This corresponds to a magnitude change of $\sim$+1.7-3.0 mag, which puts them reasonably on the giant branch. 

In the bottom plot of Figure \ref{fig:data}, we show the Galactic longitudes and latitudes for the fields in our survey as red points. We also show an extinction map in the background from \citet{Planck2013dust}. The light blue box roughly indicates the region of the bulge and the light blue point indicates the Galactic center (GC) at ($l,b$)=(0$^{\circ}$,0$^{\circ}$). Our observations have a range of Galactic longitudes that span from the center to one edge of the bulge's major axis. We also have observations from two different Galactic latitudes. However, as the bulge has a vertical metallicity gradient where the larger latitudes are generally more metal-poor \citep{Zoccali2008,Gonzalez2011,Johnson2011,Johnson2013a}, most of our observations are concentrated there. 

\subsection{High Resolution UVES Spectra}

 In this work, we made use of the radial velocities (RVs) and metallicities from COMBS I \citep{Lucey2019}, which reduced and analyzed the UVES spectra. For a complete description of the UVES spectra and reduction see Section 3.1 of \citet{Lucey2019}. In short, the UVES observations were taken in the standard RED580 setup. This setup has R$\sim$ 47,000 and wavelength coverage of 4726-6835 \AA\ with a gap (5804-5817 \AA) between the lower/blue and upper/red chips. In \citet{Lucey2019}, we reduced the data using the FLAMES-UVES workflow within the EsoReflex interface\footnote{\url{https://www.eso.org/sci/software/esoreflex/}}. We  continuum normalized, RV corrected and co-added the spectra using iSpec \citep{Blanco-Cuaresma2014}. After removing stars with signal-to-noise ratio (SNR)   < 10 $\rm{pixel^{-1}}$ in the red part of the spectrum, we are left with 35 stars that we use the RV measurements of in this work. 

In \citet{Lucey2019}, we also measured the metallicities for 26 of these stars. The metallicities were determined using the standard Fe-Excitation-Ionization balance technique through the Brussels Automatic Code for Characterizing High accUracy Spectra\footnote{\url{http://ascl.net/1605.004}} \citep[BACCHUS,][]{Masseron2016}. In short, BACCHUS uses an iterative technique to simultaneously solve for the effective temperature (\Teff), the surface gravity (\logg), microturbulence (\vmicro)  and metallicity ([Fe/H]). The \vmicro\ is solved when the Fe abundance derived from the core line intensity and equivalent width for each line are consistent. For the \teff\ and \logg\ determination as well as the final reported [Fe/H], the Fe abundance is computed using a $\rm{\chi^2}$ minimization between a synthesized spectrum and the observed spectrum for each Fe line. The \teff\ is solved when there is no correlation between the excitation potential and abundance of the line. The \logg\ is solved when there is no offset between the neutral (Fe I) and singly ionized (Fe II) line abundances. Although this process is automated, we visually inspect the line fits and validity of the solution for each star. For more details on how BACCHUS derives the stellar parameters, see Section 4 in \citet{Lucey2019}.

\subsection{Medium Resolution GIRAFFE Spectra}

\begin{figure*}
    \centering
    \includegraphics[width=\linewidth]{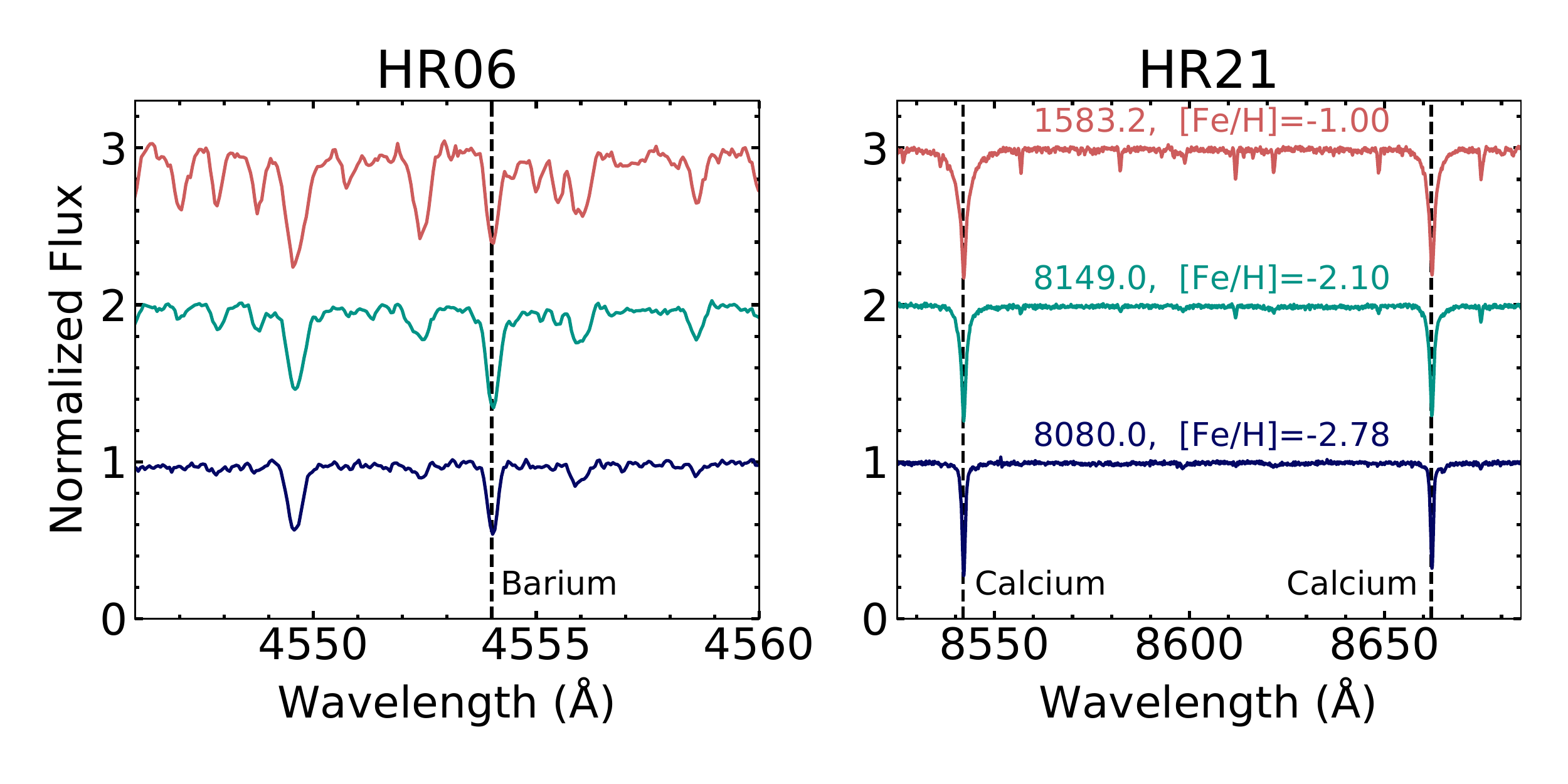}
    \caption{Partial regions of three observed spectra with varying metallicities. Specifically we show the spectra of 1583.2 (red), 8149.0 (green) and 8080.0 (dark blue). On the left is part of the HR06 spectra while on the right is part of the HR21 spectra with two of the Ca-II triplet lines shown. On the left side, we inidcate Barium line at 4554 \AA.}
    \label{fig:spec}
\end{figure*} 

\begin{figure}
    \centering
    \includegraphics[width=\columnwidth]{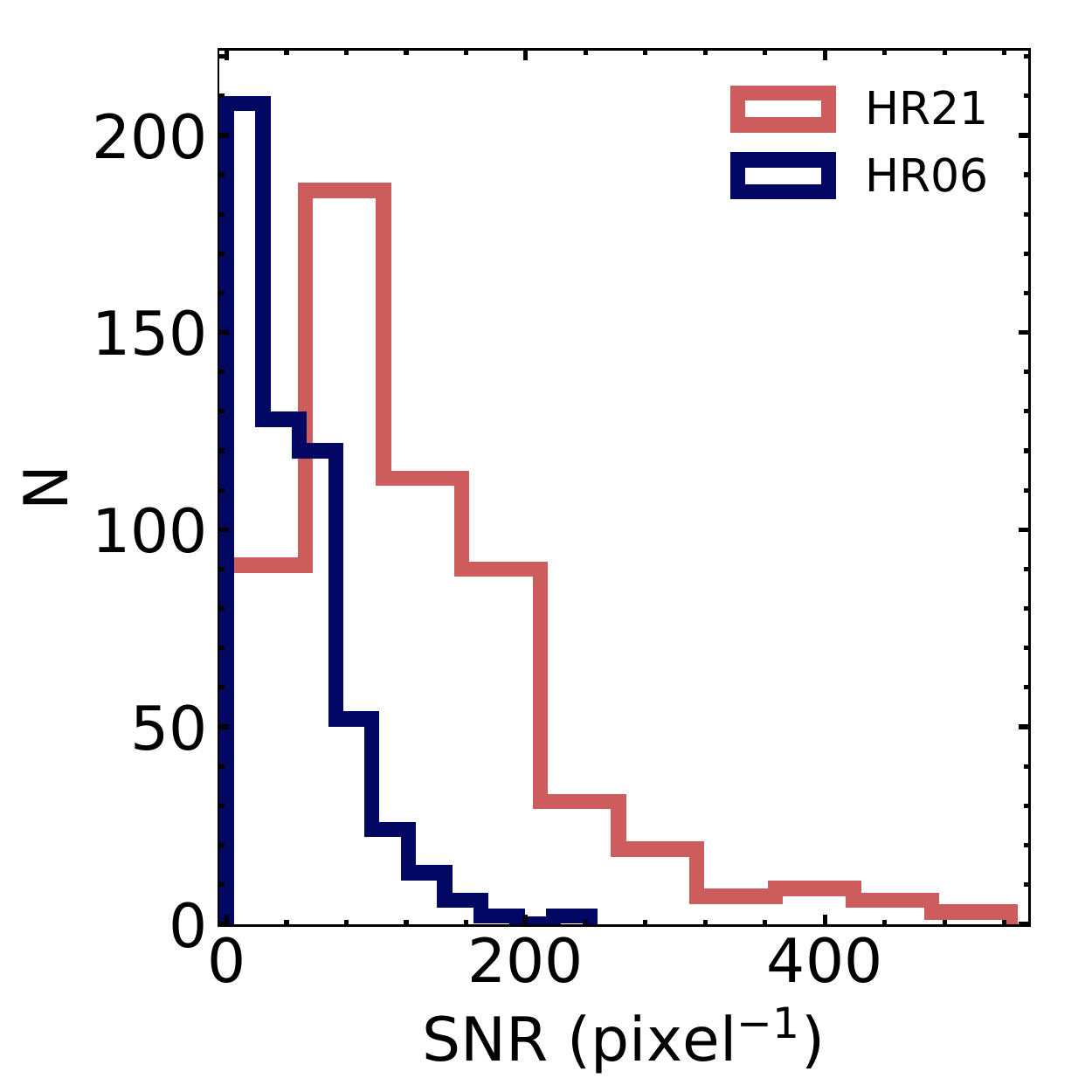}
    \caption{Distribution of the signal-to-noise ratios (SNR) per pixel for the HR06 spectra and HR21 spectra. The SNR is determined using the measured flux errors. The HR21 spectra typically have higher SNR because they are redder and are therefore less impacted by extinction.  }
    \label{fig:snr}
\end{figure} 

The GIRAFFE spectrograph in MEDUSA mode can range from medium to high resolution (R=5500-38000) with possible, although not complete, wavelength coverage from 3700-9000 \AA.  For the high resolution mode, this wavelength coverage is divided into 22 different possible setups. For this work, we use the high resolution MEDUSA HR06 and HR21 setups. The HR06 setup has R$\sim$24,300 with wavelength coverage from 4538-4759 \AA. The HR21 setup has R$\sim$18,000 with wavelength coverage 8484-9001 \AA.  We chose the HR21 setup because it contains the CaT, which provides precise radial velocities and accurate metallicity estimates \citep[e.g.,][]{Steinmetz2020a,Steinmetz2020b}. The HR06 set up is useful for deriving stellar parameters and elemental abundances because it contains many metal lines including a Barium line (4554 \AA). We show three examples of spectra with varying metallicities in Figure \ref{fig:spec}. Specifically, on the left, we show a part of the HR06 spectra with the Barium line at 4554 \AA. On the right, we show a part of the HR21 spectra with two of the CaT lines.  For more information about the FLAMES/GIRAFFE instrument we refer the reader to \citet{Pasquini2000}.

We reduced the GIRAFFE spectra using the workflow\footnote{\url{ftp://ftp.eso.org/pub/dfs/pipelines/instruments/giraffe/giraf-reflex-tutorial-1.3.pdf}} in the EsoReflex interface. We downloaded the calibration files from the ESO archive\footnote{\url{http://archive.eso.org/eso/eso_archive_main.html}} using the CalSelector tool\footnote{\url{http://www.eso.org/sci/archive/calselectorInfo.html}}. In short the workflow performs standard bias and flat-field subtraction, fiber-to-fiber corrections, wavelength calibration and extraction. We also turn on the cosmic ray cleaning feature using the package PYCOSMIC\footnote{\url{http://www.bhusemann-astro.org/?q=pycosmic}}.

In addition to the EsoReflex workflow reduction, we also perform sky subtraction. As multiple fibers per pointing observed the sky, we create a master sky spectrum for each of the pointings. We then use the IRAF function SKYTWEAK to perform the sky subtraction for the science spectra. The rest of the reduction is done using iSpec \citep{Blanco-Cuaresma2014}. We RV correct the spectra using a cross-correlation with respect to an Arcturus spectrum. As each target was observed multiple times, we then co-add the spectra of each unique target. We only add spectra whose individual SNR > 10 $\rm{pixel^{-1}}$. As the EsoReflex pipeline returns flux error estimates, we determine the SNR by dividing the flux value of each pixel by the flux error and taking the median of all the pixels. We then continuum normalize the co-added spectra using a third-order spline.  Figure \ref{fig:snr} shows the SNR values for HR06 and HR21, respectively. The HR06 spectra generally has lower SNR because the high levels of extinction toward the Galactic center preferentially remove bluer light. There are 5 stars that do not have a single observation with SNR > 10 $\rm{pixel^{-1}}$ and therefore we only report RVs for 550 stars. In addition, there are 545 stars observed with HR21 that has SNR > 10 $\rm{pixel^{-1}}$ and only 394 stars observed with HR06 that has SNR > 10 $\rm{pixel^{-1}}$.

\subsection{Parallaxes and Proper Motions from Gaia}

We use \gaia\ DR2 data in order to do full 3-D dynamical and orbit analysis for our stars. We perform a sky-crossmatch using the right ascension (RA), declination (DEC) to acquire the parallaxes, proper motions, and full covariance matrix for each of our stars. As the parallax and proper motions are highly covariant, it is essential that we include the covariances in our analysis to ensure we do not underestimate our final reported errors on the Galactic positions and velocities of our stars. Out of the 550 GIRAFFE spectra with RV measurements,  only 541 stars have a match in the \gaia\ DR2 catalog within 1 arcsecond. All 35 stars with RV measurements from the UVES spectra have a match within 1 arcsecond in the \gaia\ DR2 catalog.

\citet{Lindegren2018b} demonstrated that only using stars with renormalized unit weight error (ruwe) <1.4 is as, if not more, effective at removing problematic astrometry than the quality cuts recommended by \citet{Gaiasummary2018,Lindegren2018a,Arenou2018}. Therefore, similar to recent literature \citep[e.g.,][]{Anders2019,Lucey2020}, we only use $Gaia$ DR2 data with ruwe<1.4. This leaves us with a total of 523 stars, including 31 stars with UVES spectra, with which we can perform 3D dynamical and orbit analysis.

\section{Metallicity Estimates from Ca-II Triplet} \label{sec:metallicities}

\begin{figure}
    \centering
    \includegraphics[width=\columnwidth]{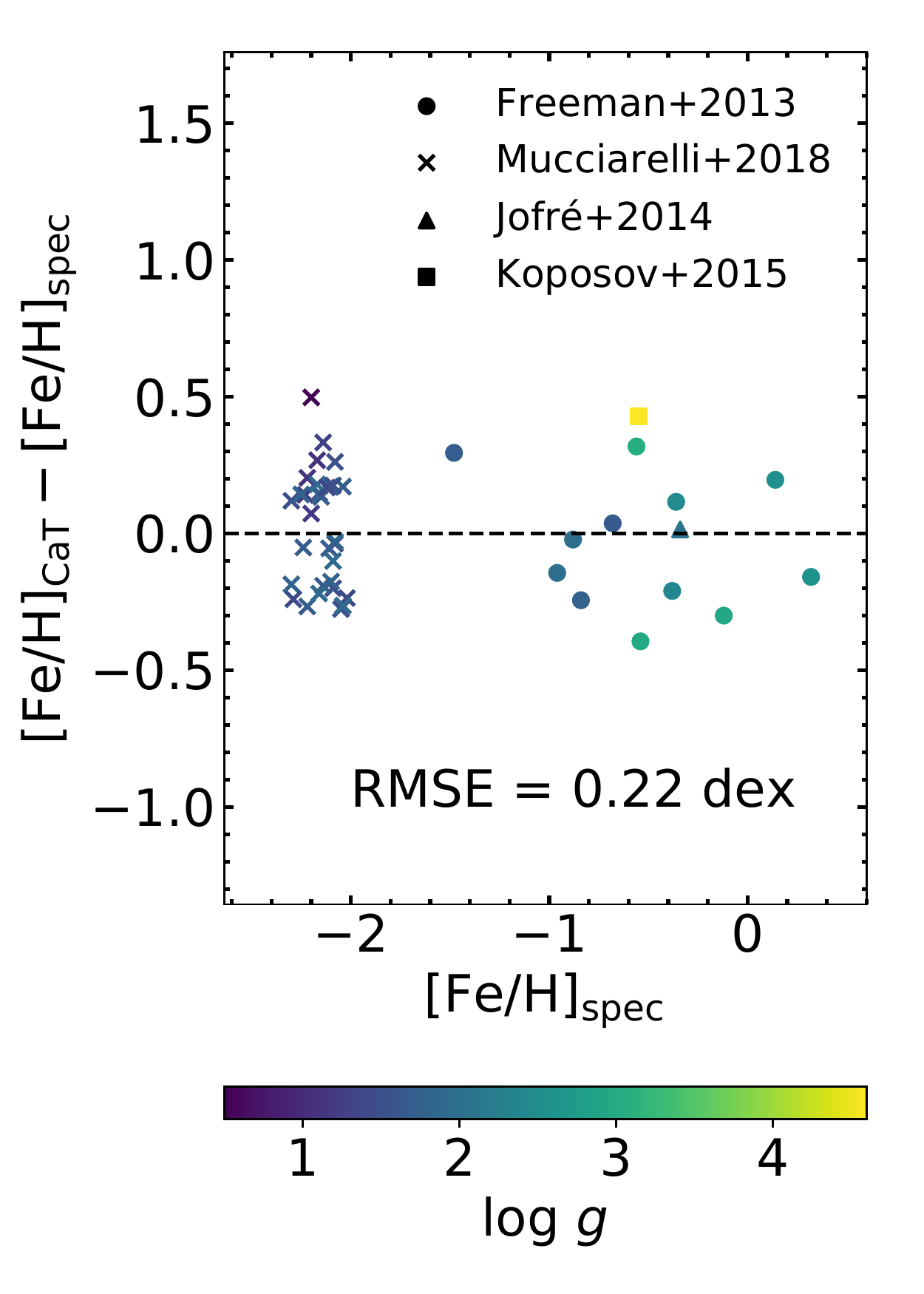}
    \caption{Comparison of metallicity estimates from the Ca-II triplet to the values derived from Fe lines using full spectroscopic analysis for the entire validation sample of 45 stars. The spectroscopic values for the validation sample are taken from 4 different studies \citep{Freeman2013,Mucciarelli2018,Jofre2014,Koposov2015}, indicated by the marker shape. We quantify the precision of the estimate using the root-mean-square error (RMSE), which equals 0.22 dex. The points are colored by the surface gravity (\logg).}
    \label{fig:met}
\end{figure} 
\begin{figure}
    \centering
    \includegraphics[width=\columnwidth]{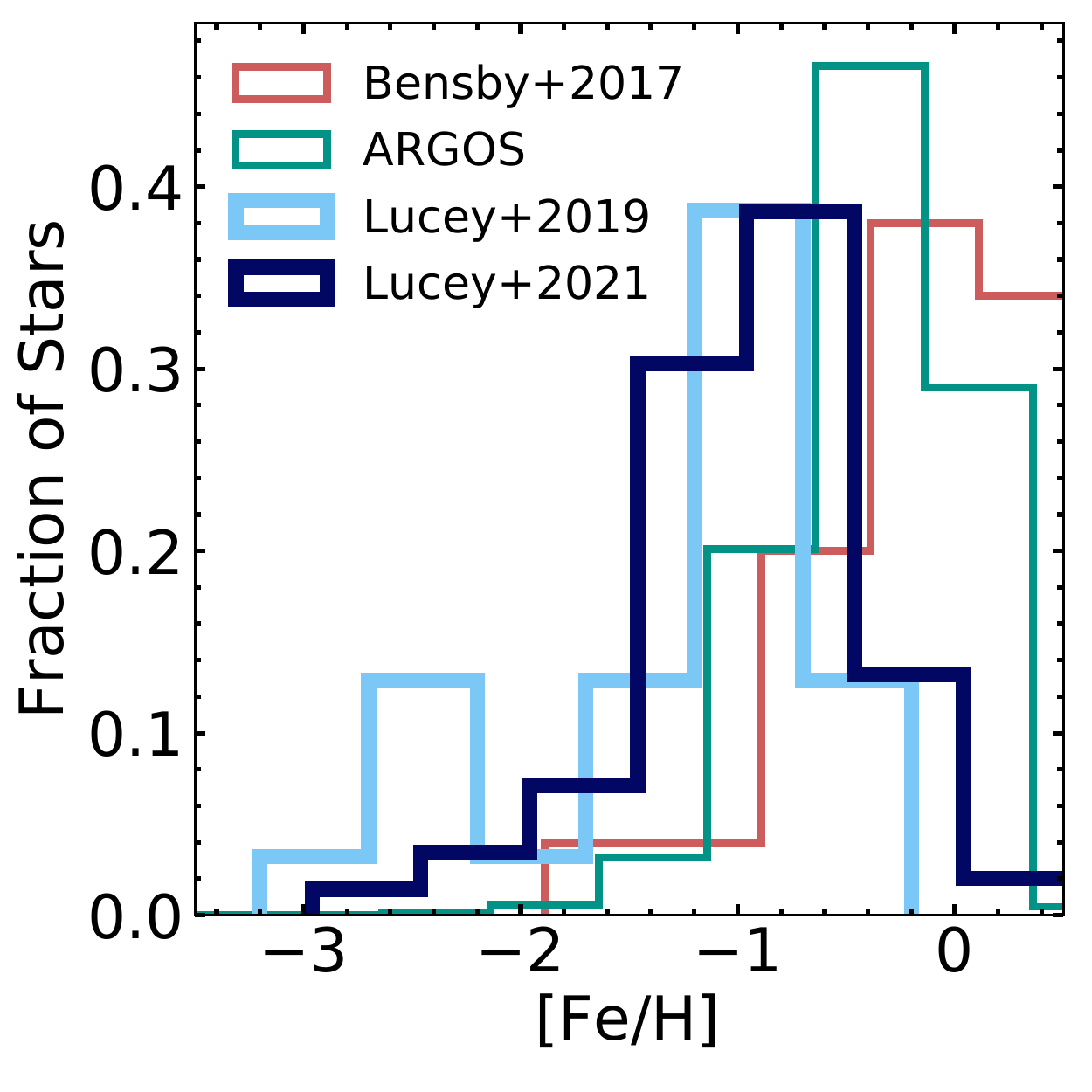}
    \caption{Derived metallicity distribution function for the 473 GIRAFFE spectra compared to the results for the 26 UVES spectra \citep{Lucey2019}, the ARGOS survey \citep{Freeman2013} and \citet{Bensby2017}. We have successfully targeted the metal-poor tail of the bulge metallicity distribution function. The distribution for the GIRAFFE spectra are not as metal-poor as the UVES spectra, which is expected given that the most promising metal-poor targets were prioritized for the higher resolution data. }
    \label{fig:mdf}
\end{figure} 

The CaT is frequently used to determine metallicities from moderate resolution spectra \citep[e.g.,][]{Armandroff1988,Olszewski1991,Armandroff1991,Starkenburg2010,Li2017}. It has been shown that the equivalent widths (EW) of the CaT can provide accurate metallicity estimates within $\sim$0.1 dex, irrespective of age effects \citep{Cole2004}. However, early work demonstrated it is essential to account for the sensitivity to surface gravity (\logg) \citep{Spinrad1969,Spinrad1971,Cohen1978,Jones1984}. The most common method to accomplish this is to use the absolute magnitude of the star in the calibration. Unfortunately, determining the absolute magnitude for our bulge stars is extremely difficult given the high and varying levels of extinction along with the large uncertainties on the distance estimates. 

In this work, we develop a new method to estimate the metallicity from the CaT for the GIRAFFE spectra.  As some of the stars observed in this program were also observed in the ARGOS survey \citep{Freeman2013}, we use those to calibrate our metallicities. We also supplement these data with other metal-poor samples from the literature that have spectroscopic metallicities and have been observed with the GIRAFFE HR21 setup. These samples are of NGC 5824 \citep{Mucciarelli2018}, Reticulum 2 \citep{Koposov2015}, and a number of $Gaia$ benchmark stars \citep{Jofre2014}. These data were downloaded from the ESO archive and reduced using the same methods as our program spectra. Consistent with previous work \citep[e.g.,][]{Armandroff1991,Battaglia2008,Starkenburg2010}, we use only the two strongest CaT lines, 8542 \AA\ and 8662 \AA, whose equivalent widths can be measured more accurately. We fit a Voigt profile, which is a combination of a Lorentzian and Gaussian profile, to these lines and define the EW as the integral of the fitted function. The wings of the CaT line have proven to be powerful for constraining the \logg\, in addition to the metallicity, of giant stars \citep{Jones1984,Freeman2013,Arentsen2020b}. These works indicate that there should be both \logg\ and metallicity information embedded in the line profiles of the CaT. As such, we include the Voigt profile fit parameters in our calibration in order to calibrate out the impact of \logg\ on our metallicity determination. In this way, we are essentially using the line profile information opposed to reducing this to a single number, the EW, as in previous work. However, it is important to note that we do not determine the \logg. We merely are using the line profile information in our calibration to account for the effects of \logg\ on the EW. We perform a regression where the input parameters are the mean amplitude of the Lorentzian components ($A_{Lorentz}$), the mean full-width-half-maximum (FWHM) of the Lorentzian components ($\sigma_{Lorentz}$), the mean FWHM of the Gaussian components ($\sigma_{Gauss}$), and the sum of the EWs of the two lines ($EW_{\Sigma}$). We also input the square of these parameters in order to allow for a non-linear, second-order relation. For completeness, we also try higher-order relations but found that the increase in precision was negligible. The final relation we derive is:

\begin{equation}
\begin{split}
    \rm{[Fe/H]} = & -0.99 -0.80EW_{\Sigma} +3.46\sigma_{Lorentz} +7.12\sigma_{Gauss} \\  &+10.07A_{Lorentz}+0.08EW_{\Sigma}^2-0.49\sigma_{Lorentz}^2 \\
    &-42.09\sigma_{Gauss}^2+7.72A_{Lorentz}^2
\end{split}
\end{equation}

We use 70\% of our sample with known metallicities to calibrate the model and the remaining 30\% to validate. We show the comparison between the literature metallicities to the metallicities we derive for our validation sample in Figure \ref{fig:met}. Although we are unable to find a reference star in the literature across all metallicities, we have no reason to expect that the relation does not interpolate well or is unable to extrapolate slightly. We are able to recreate the metallicities to a precision of 0.22 dex over a wide range of \logg. It is important to note that the precision is not a function of \logg\ or metallicity. As our method is data-driven, the precision is limited by the precision of the training data with which we calibrate our method. Our calibration sample generally has metallicity uncertainties between $\sim$0.05-0.15 dex \citep{Freeman2013,Jofre2014,Koposov2015,Mucciarelli2018}. It is also possible that there are systematic offsets in the metallicity scale between the 4 bodies of work from which we source our calibration sample. Offsets in metallicity between bodies of work is typical and can be as high as $\sim$0.2 dex depending on the lines, atomic data and methods used \citep{Yong2013,Bensby2014,Lucey2019}. It is likely that the offsets between the literature values, from which we derive our calibration sample, also decreases our precision. Previous work on the CaT metallicity calibration has achieved a precision of $\sim$0.1-0.2 dex  \citep{Battaglia2008,Carrera2013}. However, these methods rely on an accurate estimate of the luminosity to account for the impact of the \logg\ on the EW. In this work, we achieve a precision of 0.22 dex, which is competitive to previous studies. In total, our method provides a unique way to derive metallicities from the CaT that achieves similar precision to previous results without depending on an estimate of the luminosity. 

We apply this calibration to our entire sample of 492 stars that have a HR21 spectrum with SNR > 10 $\rm{pixel^{-1}}$ and a match in \gaia\ DR2 with ruwe<1.4. The majority of our stars are giant stars with \logg\ > 3.5 dex (see Figure \ref{fig:data}). This is consistent with our calibration sample which also primarily consists of giant stars. However, we also likely have some sub-giant stars in our sample (see Figure \ref{fig:data}). Therefore we include stars with \logg\ as high as 4.5 dex in our calibration sample. In order to avoid extreme extrapolation, we only keep stars with -3 dex $<\rm{[Fe/H]_{CaT}}<0.5$ dex given that our calibration sample has -2.74 dex $\leq$[Fe/H]$\leq$ 0.32 dex. This leaves us with 473 out of 492 GIRAFFE spectra with metallicity estimates from the CaT. We show the final metallicity distribution of our sample in Figure \ref{fig:mdf} along with a comparison to the results for the UVES spectra from \citet{Lucey2019}, the ARGOS survey \citep{Freeman2013}, and a survey of bulge micro-lensed dwarf stars \citep{Bensby2017}. From Figure \ref{fig:mdf}, it is clear we have successfully targeted metal-poor stars compared to the bulge surveys which did not specifically target metal-poor stars \citep[ARGOS;][]{Freeman2013,Bensby2017}. However, we also do not have as large of a low metallicity tail ([Fe/H] <-2) as seen in \citet{Lucey2019}. This is as expected because the most promising metal-poor stars were prioritized to be observed with the higher resolution setup and were therefore included in the UVES sample.

\section{Dynamical Analysis} \label{sec:dynamics}
One of the main goals of our work is to determine if the metal-poor stars that are currently in the bulge are of the bulge. However, this is difficult given that the majority of our data have \gaia\ DR2 fractional parallax uncertainties > 50\%. With high parallax uncertainties, probabilistic Bayesian inference affords a useful approach for determining stellar distances \citep[e.g.,][]{Bailer-Jones2018} and subsequently their orbital properties.

\subsection{Galactic Positions and Velocities}
\label{sec:positions}

\begin{figure}
    \centering
    \includegraphics[width=\columnwidth]{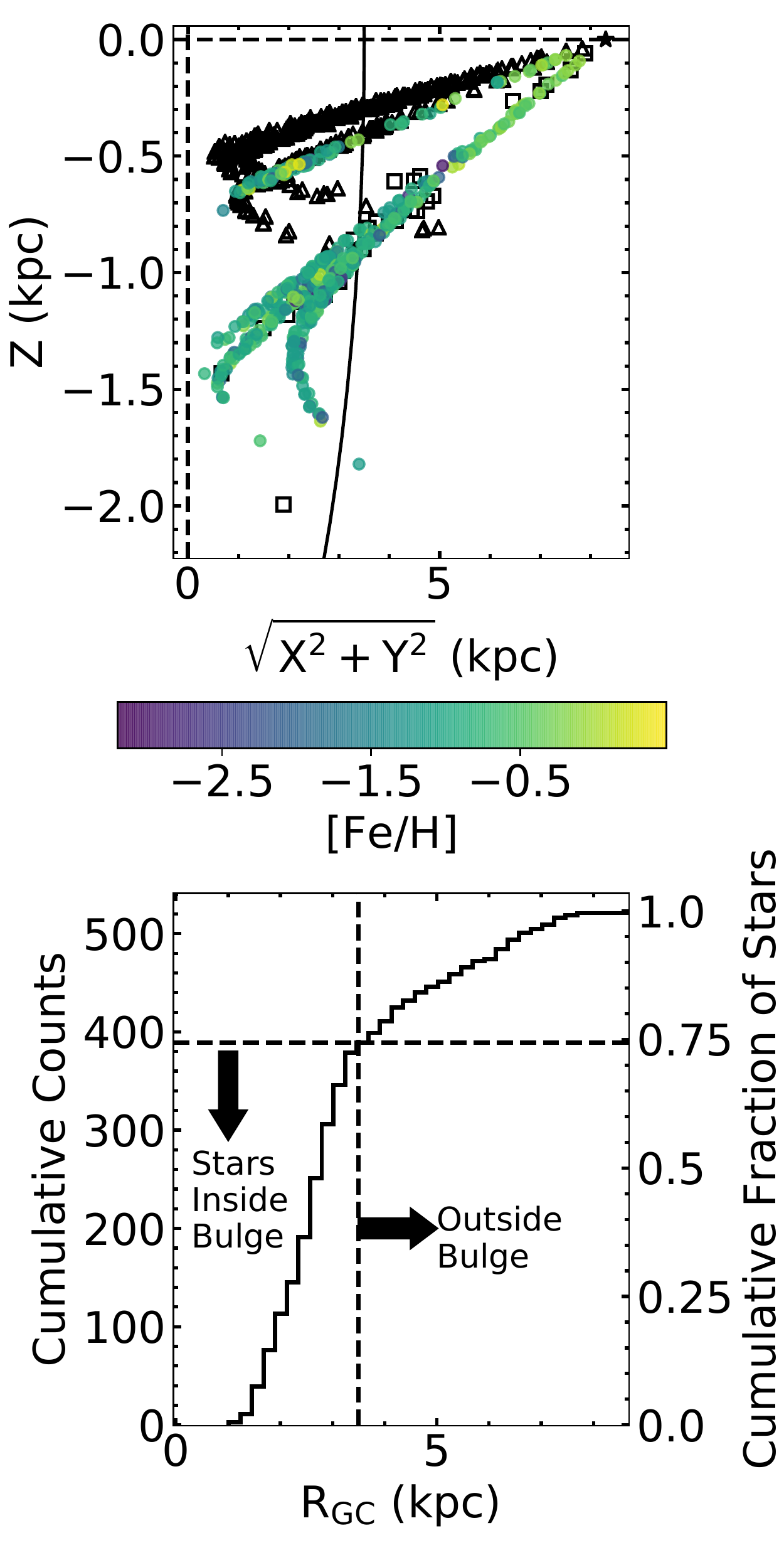}
    \caption{The top plot shows the positions of our observed stars with respect to the Galactic center (0,0) colored by metallicity. We also show the GIBS \citep{Gonzalez2015} and EMBLA \citep{Howes2016} samples in black open triangles and open squares, respectively. The Sun is shown as a black star at (8.3,0) kpc. We also show the outline of what we define as the bulge at a distance of 3.5 kpc from the Galactic center as a solid black line. We have some contamination in our sample from metal-rich disk stars along the line-of-sight towards the bulge. In the bottom plot we show the cumulative distribution of the distance from the Galactic center ($R_{GC}=\sqrt{X^2+Y^2+Z^2}$) where the vertical dashed line corresponds to a distance of 3.5 kpc, which 73\% of the sample (381 stars) lies within.}
    \label{fig:dists}
\end{figure}  

To determine the Galactic positions and velocities, we use a Markov Chain Monte Carlo (MCMC) simulation and  Bayesian inference, which allows us to incorporate prior information on Galactic structure and the covariances between the positions and velocities. We first infer the distance and proper motions using the parallax and proper motion data from \gaia\ DR2 along with the covariance matrix. Although the proper motions are measured by \gaia\, it is necessary to reinfer them with the distance in the context of the prior and covariances. Our prior on the proper motions is flat, while our prior on the Galactic distance is based on the \gaia\ DR2 mock catalog from \citet{Rybizki2018}. Specifically, we use the star counts as a function of distance, which changes as a function of line-of-sight, as an unnormalized probability distribution function. With the use of an MCMC simulation, it is not necessary to normalize this distribution. This is different from the \citet{Bailer-Jones2018} catalog, which uses an exponentially decreasing prior with a scale length that varies as a function of line-of-sight. An exponentially decreasing model does not accurately describe the distribution of stars when looking towards the Galactic center. Therefore, using the mock catalog provides a more realistic prior. Nonetheless, when we compare our results to the catalog from \citet{Bailer-Jones2018} we find the results are generally consistent. Only three stars have distances that are inconsistent with the \citet{Bailer-Jones2018} results. These stars all have negative parallaxes and distances of $\sim$13 kpc in the \citet{Bailer-Jones2018} catalog, which would put them all outside of the bulge. Only one of these stars has a shorter distance in our catalog and is determined to be currently within the bulge.

We then use the RA, DEC, and measured RV to convert the proper motions and distances into 3D Galactic positions and velocities. To do this, we sample normal distributions for the RA, DEC and RV that are centered on the measured values with widths equivalent to the measured errors. We create as many samples as the length of the MCMC chain. We then combine these samples with the MCMC chain to calculate the 3D Galactic positions and velocities with the covariances propagated through.

We show the Galactocentric distribution of the 523 stars in Figure \ref{fig:dists}. The top panel shows the cylindrical Galactocentric positions ($R=\sqrt{X^2+Y^2},Z$) colored by the metallicities. We also show literature bulge studies from the GIBS survey \citep{Gonzalez2015} and the EMBLA survey \citep{Howes2015} in black for comparison. We show the position of the Sun as a black star at (8.3,0) kpc 
\citep{Reid2014}. We also show the edge of what we consider the bulge as a black line, which corresponds to a distance of 3.5 kpc from the Galactic center and is consistent with what is typically used in the literature \citep[e.g.,][]{Ness2013b,Arentsen2020,Kunder2020}. Our sample clearly has some contamination from disk stars that are along the line-of-sight towards the bulge. These stars are typically more metal-rich than the stars that are within or close to within the bulge. This contamination is typical of bulge surveys, including the EMBLA \citep{Howes2015} and GIBS \citep{Gonzalez2015} surveys. In the bottom panels we show the cumulative distribution of the distance from the Galactic center for our sample. The vertical dashed line indicates a distance of 3.5 kpc. The dashed horizontal line corresponds to the number of stars within 3.5 kpc (381) on the left y-axis  and the fraction of stars that are within 3.5 kpc (0.73) on the right y-axis. Therefore, 73\% of our sample, or 381 stars, are currently within the bulge.

\subsection{Orbital Properties}
\label{sec:orbits}
We aim to determine whether the metal-poor stars currently in the bulge are confined to or merely passing through the bulge. To this end, we calculate the orbits of all stars in the sample. We do this using the GALPY\footnote{http://github.com/jobovy/galpy} package and the 2014 MW potential \citep{Bovy2015}. However, this potential is axisymmetric and does not contain a bar. Since all of our stars are near to the Galactic center it is essential that we add a bar to this potential. Therefore, we add a Dehnen bar potential \citep{Dehnen2000} generalized to 3D following \citet{Monari2016}:
 
\begin{equation}
\begin{split}
     \Phi(R,z,\phi) = &A_f\rm{cos}(2(\phi-\phi_b-\Omega_bt))\left( \frac{R}{r} \right)^2 \\ &\times \begin{cases} -(R_b/r)^3, & \text{if } r\geq R_b \\
     (R_b/r)^3-2, & \text{if } r \leq R_b
     \end{cases}
\end{split}
\end{equation}
where $r=\sqrt{R^2+z^2}$ is the spherical radius, $R_b$ is the bar radius, $\Omega_b$ is the rotation speed of the bar, $\phi_b$ is the bar angle and $A_f$ is the bar strength. The bar strength is defined as $\alpha$, where $\alpha=3(A_f/v_0^2)(R_b/r_0)^3$, $v_0$ is the local circular speed and $r_0$ is the Sun's distance from the Galactic center. This potential is included in the GALPY package. We use measured MW parameters to intialize the bar potential. Specifically, we use $\phi_b$= 27$^{\circ}$ \citep{Wegg2013}, $\alpha$=0.01 \citep{Monari2016}, $R_b$=5 kpc \citep{Wegg2015}, and $\Omega_b$=39 km/s $\rm{kpc^{-1}}$ \citep{Portail2017}\footnote{We also performed the analysis using parameters for a shorter, faster bar. Specifically, we used $\phi_b$= 25$^{\circ}$ \citep{Dehnen2000}, $\alpha$=0.01 \citep{Monari2016}, $R_b$=3.5 kpc \citep{Dehnen2000}, and $\Omega_b$=52.2 km/s $\rm{kpc^{-1}}$ \citep{Dehnen2000}. Using these parameters only decreases the number of stars that stay confined to the bulge by $\sim$5\% and does not impact our conclusions. }.
 
For each star we pick 1000 random points from the MCMC chain of positions and velocities. We then initialize 1000 different orbits at those points in order to propagate the errors and covariances through to the orbital properties. We integrate all of the orbits for 1 Gyr. We report the orbital properties (ecccentricity, apocenter, pericenter, $z_{max}$) as the median of those 1000 orbits and the asymmetric errors as 1$\sigma$. In addition, we report the probability that a star stays confined to the bulge (P(conf.)) as the number of orbits out of the 1000 that have apocenter < 3.5 kpc divided by 1000.

\section{Do Metal-Poor Stars in the Bulge Stay in the Bulge?}
\label{sec:confined}

\begin{figure}
    \centering
    \includegraphics[width=\columnwidth]{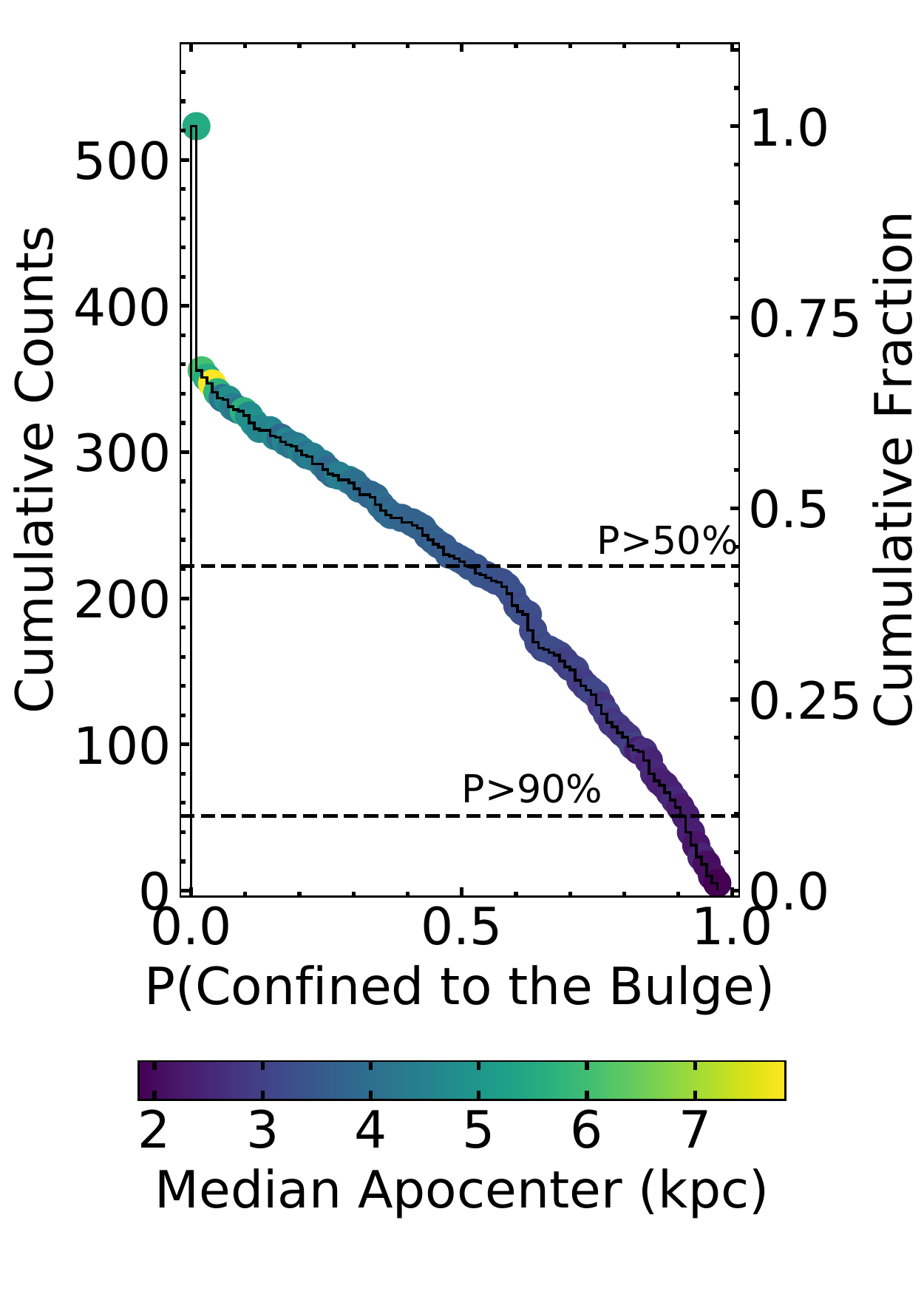}
    \caption{The distributions of probabilities that the stars stay confined to the bulge, which we define as within 3.5 kpc from the Galactic center. The points are colored by the median apocenter at that probability. The dashed lines correspond to the number of stars with probability > 50\% and >90\%, which are  $\sim$ 43\% and $\sim$ 10\% of the sample, respectively. }
    \label{fig:probs}
\end{figure}  

\begin{figure}
    \centering
    \includegraphics[width=\columnwidth]{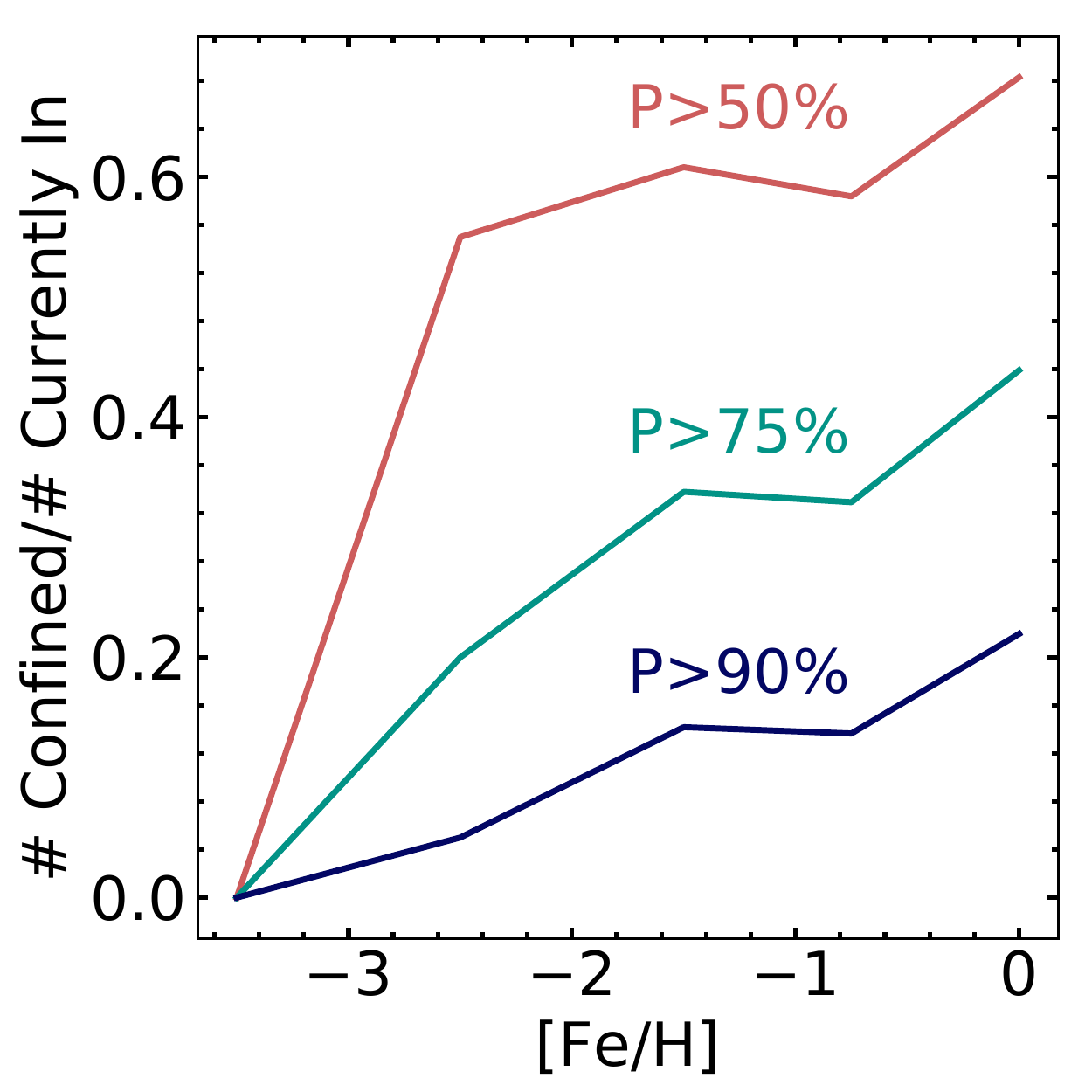}
    \caption{The fraction of stars that are currently in the bulge that have a >50\% (red), >75\% (green) and >90\% (dark blue) probability of staying within 3.5 kpc of the Galactic center as a function of metallicity. }
    \label{fig:frac}
\end{figure} 

The first step toward determining the origins of the metal-poor bulge stars is to separate the confined bulge stars from the halo interlopers.  In this section, we use the measured probabilities of being confined to the bulge, which are defined in Section \ref{sec:orbits}, to determine the rate at which our sample is contaminated by halo stars.

In Figure \ref{fig:probs}, we show the reverse cumulative distribution of the probabilities that the stars are confined to the bulge. We color the line by the median apocenter of stars with that probability to demonstrate that P(conf.)$\approx$ 50\% corresponds to a median apocenter of $\sim$3.5 kpc. The dashed lines correspond to the number of stars with P(conf.)> 50\% (223 stars or $\sim$42\% of the sample) and P(conf.)> 90\% (54 stars or $\sim$10\% of the sample).  Based off the derived Galactic positions, we determined that 73\% or 381 stars are currently within the bulge (see Section \ref{sec:positions}). Of these 381, only 223, or 59\%, have P(conf.)> 50\%. Therefore, almost half of the stars in our sample are likely halo interlopers. However, it is possible that many of these stars that do not stay confined to the bulge could be metal-weak thick disk stars or bulge stars that have apocenters only slightly larger than 3.5 kpc. Although, most of the stars that do not stay confined have eccentricity >0.6 and apocenter > 6 kpc, indicating that they are most likely halo stars.

We also find that the percentage of stars that stay confined to the bulge decreases with decreasing metallicity. In Figure \ref{fig:frac}, we show the fraction of stars that will stay in the bulge with various probabilities over the number of stars currently in the bulge as a function of metallicity. However, the number of stars in our sample also decreases with decreasing metallicity for [Fe/H] <-1 dex. For example, there is only 1 star with [Fe/H] < -3 dex in our sample that is currently within the bulge \citep{Lucey2019}. This star has a P(conf.) = 0\%. There are 21 stars in our sample with -3 dex $\leq$ [Fe/H] < -2 dex that are currently in the bulge. Only 11 of these stars have P(conf.) > 50\%. However, this drops to 4 stars when we restrict to stars with P(conf.) > 75\%. 

These results demonstrate the importance of performing orbit analysis to remove the contamination when studying metal-poor bulge stars, especially for stars with [Fe/H]< -2 dex. Previous and future studies of the metal-poor star in the Galactic bulge may have different selection functions, which may result in differing rates of contamination by halo interlopers. For example, \citet{Kunder2020} found that only 25\% of their sample of RR Lyrae stars had apocenters> 3.5 kpc. However, we note that the kinematic results, specifically the Galactocentric line-of-sight velocity distributions as a function of Galactic longitude, for studies which did not target RR Lyrae stars \citep[e.g.,][]{Ness2013b,Arentsen2020} show results similar to ours when we do not remove the contamination. This may indicate similar rates of contamination with halo interlopers in these studies. Furthermore, the EMBLA survey estimates that roughly 50\% of their 23 very metal-poor stars were confined to the bulge \citep{Howes2015}, which is consistent with our results.

\section{Properties of Confined Metal-Poor Bulge Stars}
\label{sec:properties}
Now that we can separate the halo interlopers from the confined metal-poor bulge stars, we have the opportunity to study this unique population. With our data we can provide new insights on the metal-poor tail of the bulge MDF and the kinematics of these stars, which will lead to new constraints on the origins of confined metal-poor bulge stars and on the formation history of the central region of our Galaxy.

\subsection{Metallicity Distribution Function}
\label{sec:conf_met}

\begin{figure}
    \centering
    \includegraphics[width=\columnwidth]{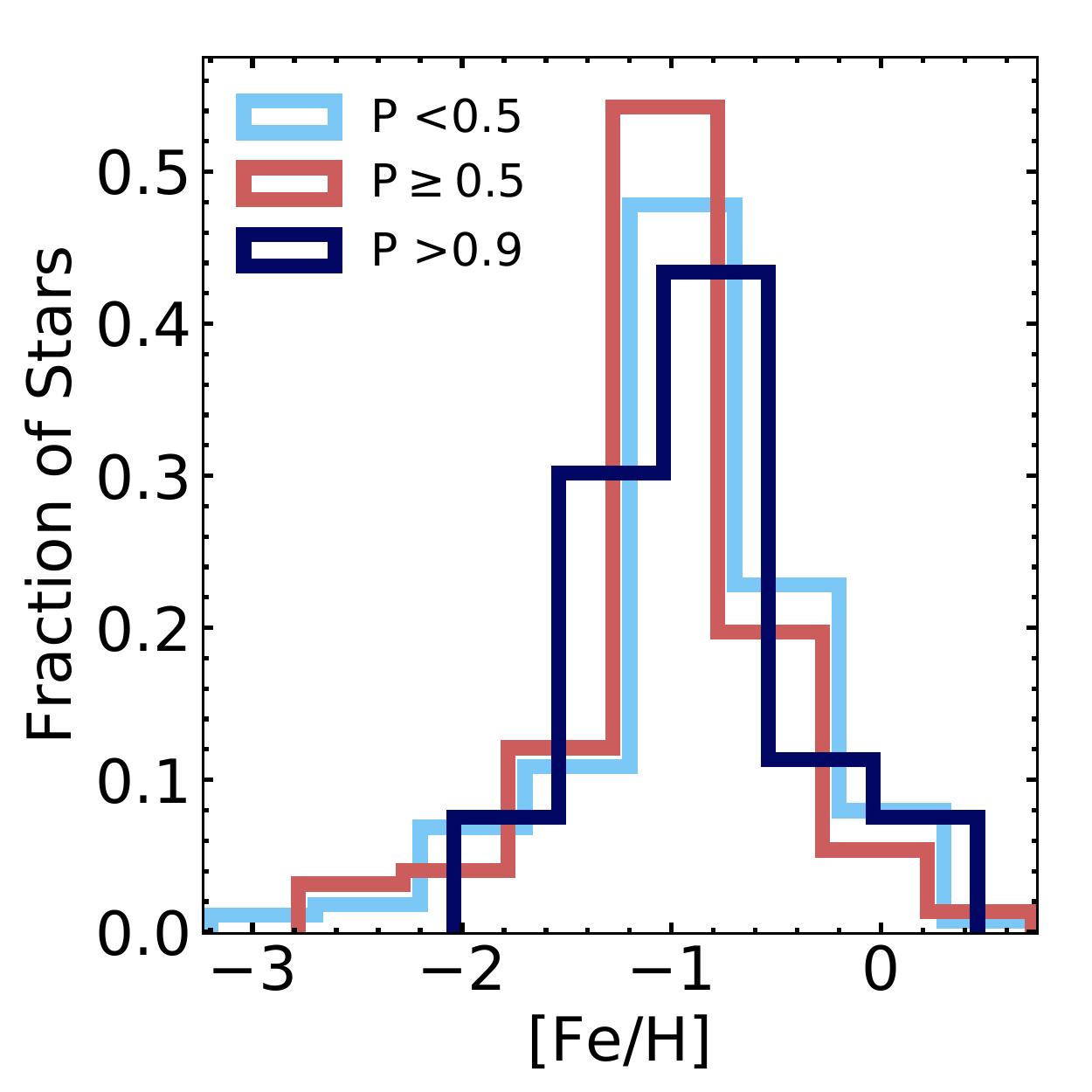}
    \caption{The metallicity distribution function of stars with different probabilities of staying confined to the bulge.   }
    \label{fig:mdf_conf}
\end{figure}
 
The MDF can provide critical information about the history of this unique metal-poor population. However, our results are heavily influenced by the metallicity selection method described in Section \ref{sec:data}. For example, the SkyMapper photometry, which is used for target selection, may be biased against selecting CEMP stars \citep{Starkenburg2017b,DaCosta2019}. If the majority of confined bulge stars with [Fe/H] <-2 dex are CEMP stars, it is possible that we would not have observed these stars. Despite this, the MDF as a function of confinement probability shows a clear trend. In Figure \ref{fig:mdf_conf}, we show the MDFs for three different cuts in the probability of confinement. In light blue, we show the stars with P(conf.) < 50\%, which are likely to be mostly halo stars and metal-weak thick disk contamination as we do not constrain the stars to be currently within 3.5 kpc of the Galactic center. In red, we show stars with P(conf.) $\geq$ 50\% and in dark blue we show stars with P(conf.) > 90\%.  As we make the cut in probability of confinement more stringent, we see the most metal-poor tail of the distribution disappears. It is important to consider that we have a small number of stars at the most metal-poor end so it is difficult to draw strong conclusions from the disappearance of this tail. Nonetheless, it is interesting to note that the metal-weak thick disk metallicity distribution is thought not to go below [Fe/H]$\approx$-1.8 dex \citep{Beers2014,Carollo2019}, which is consistent with the lowest metallicity observed for the population with P(conf.) > 90\% (-2.04 dex), indicating that these two populations may have similar origins.  However, recent results by \citet{Sestito2020} argue that the metal-weak thick disk extends to [Fe/H] <-2.5 dex. It is difficult to further compare the MDF of our stars to the thick disk because of our complicated selection function from the photometric metallicity targeting.

\subsection{Kinematics}
The kinematics of our stars can also inform us about the origins of the metal-poor bulge population. One of the main open questions about this population is whether they participate in the B/P bulge structure or if they are more consistent with a classical bulge population. In this section, we aim to answer this question and gather new insights on the history of this population.

To do this, we compare our observed kinematics to what is expected from simulations. Specifically, we use the star-forming simulation presented in \citet{Cole2014} and \citet{Ness2014}. In short, this simulation forms a disk galaxy through gas cooling and settling into a disk, which triggers continuous star formation. A bar forms in the model after $\sim$3.2 Gyr and continues to grow. By 10 Gyr, a B/P bulge has formed.  Since the bar in this model is only 3 kpc long, we multiply the spatial coordinates by 1.7 to match the MW, which has a bar measured to be 5 kpc long \citep{Wegg2015}. In addition, we multiply the velocities by 0.48, which is consistent with \citet{Ness2014} and \citet{Debattista2017}, which also use this simulation. We also rotate the model to match the position of the bar with respect to the Sun, which is at an angle of 27$^{\circ}$ from the line-of-sight to the center of the Galaxy \citep{Wegg2013}. We choose to only use stars from the simulation with the same line-of-sight towards the Galactic center as our observations and that are within 3.5 kpc of the Galactic center in order to be consistent with our observations.

\subsubsection{Line-of-Sight Velocities}

\begin{figure*}
    \centering
    \includegraphics[width=\linewidth]{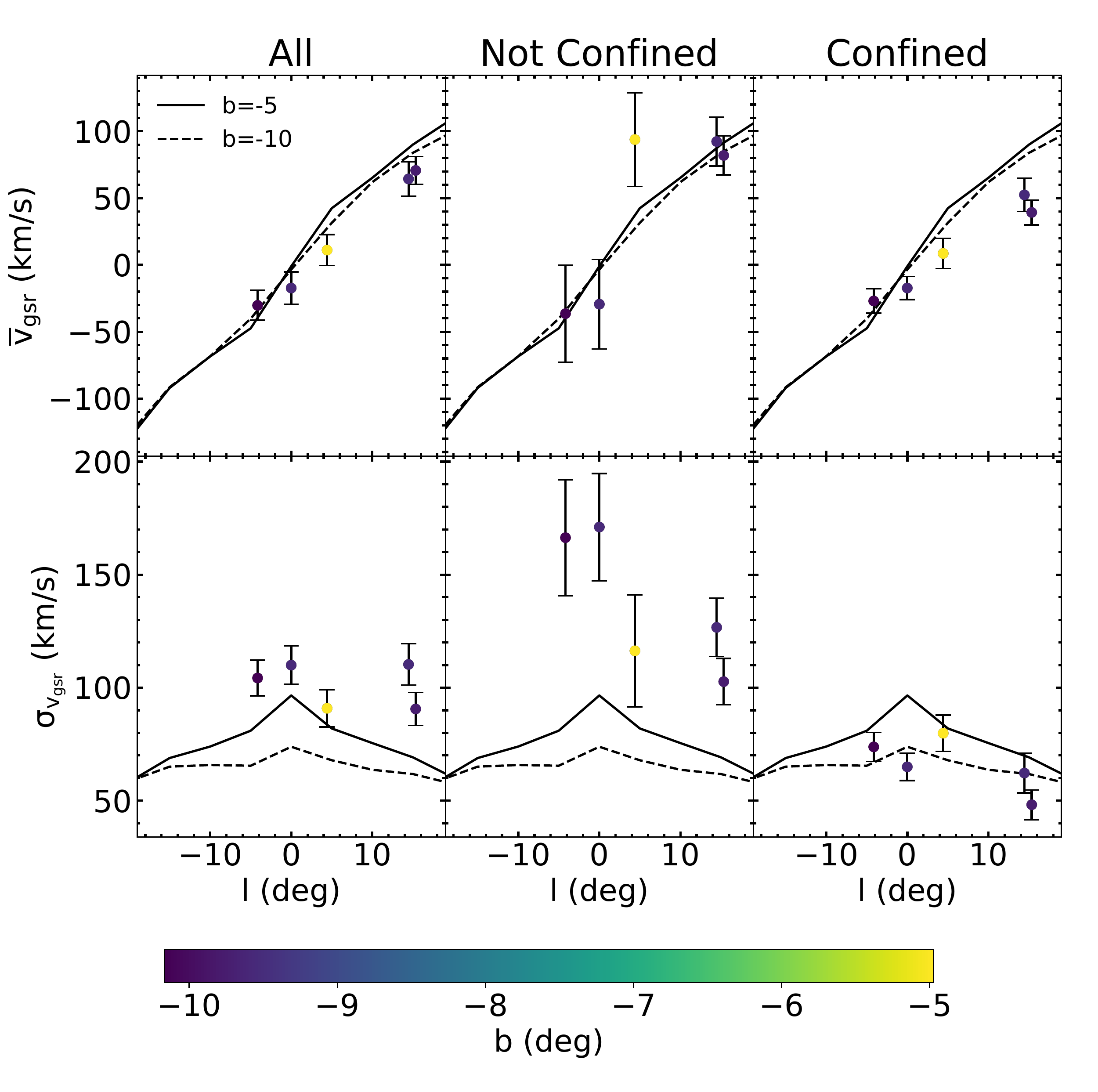}
    \caption{The mean and standard deviation of the Galactocentric line-of-sight velocities ($v_{gsr}$) as a function of Galactic longitude ($l$). The points are colored by the Galactic latitude ($b$). The error bars on the mean are $\sigma/\sqrt{N}$ where $\sigma$ is the standard deviation and $N$ is the number of stars. The error bars on the standard deviation are $\sigma/\sqrt{2N}$. In the left panel, we show results for all stars in our sample that are currently within 3.5 kpc of the galactic center. In the middle panel, we show stars with a probability of being confined to the bulge < 50\% and in the right panel, we show only stars with a probability of being confined $\geq$ 50\%. We also show results from the simulation of a B/P bulge  presented in \citet{Cole2014} and \citet{Ness2014} (black solid and dashed lines). These lines are created only using stars that formed within the first Gyr of star formation.    }
    \label{fig:los}
\end{figure*} 

Often in bulge literature, RVs are used over full 3-D motion because the proper motions and distances are poorly constrained or not measured at all. In this work, the measured RVs are considerably more precise than the 3D velocities, which depend on the distance estimate. Therefore, they can be used to provide a detailed view of bulge dynamics and an accurate comparison to the literature.  However, to understand them in a Galactic context, we first need to convert them from a heliocentric rest frame to a Galactocentric one. We convert the radial velocities to Galactic Standard of Rest ($v_{gsr}$) assuming the local standard of rest velocity at the Sun to be 220 km/s \citep{Kerr1986,Bovy2012c}, which is consistent with the Galactic potential used to calculate the orbits in Section \ref{sec:orbits} \citep{Bovy2015}. We also assume the Sun's peculiar velocity to be 17.1 km/s in the direction ($l$,$b$) = (58$^{\circ}$, 22$^{\circ}$) \citep{Coskunoglu2011}. Recent estimates of the Sun's peculiar velocity can differ by up to $\sim$3 km/s \citep{Bland-Hawthorn2016}. Therefore, adopting different values only has a small impact on our results and does not impact our conclusions.  With these values, the Galactocentric line-of-sight velocity in terms of the heliocentric radial velocity ($v_{hc}$) is then:
\begin{equation}
\begin{split}
v_{gsr} = &v_{hc} +220[\rm{sin}(\textit{l})\rm{cos}(\textit{b})] +17.1[\rm{sin}(\textit{b})\rm{sin}(22) + \\ &\rm{cos}(\textit{b})\rm{cos}(22)\rm{cos}(\textit{l}-58)]
\end{split}
\end{equation}
where $v_{hc}$ is in km/s and angles ($l$,$b$) are in degrees. 

We present the mean and standard deviation of the Galactocentric line-of-sight velocities ($v_{gsr}$) as a function of Galactic longitude ($l$) in Figure \ref{fig:los} where the points are colored by the Galactic latitude ($b$). In the left panel, we show all 523 stars. In the middle panel, we show only stars with a P(conf.) < 50\% (halo interlopers) and in the right panel we show stars with a P(conf.) $\geq$ 50\%. For comparison, we also show results from the simulation. We choose to use only stars that form within the first Gyr of the simulation as we expect these stars will be most similar to the metal-poor stars.

Stars that do stay confined have a different velocity distribution than the halo interlopers (unconfined stars). For example, the halo interlopers have a steeper slope with the Galactic longitude than the confined stars, which is indicative of faster rotation. This is especially interesting given that we expect the opposite, i.e., that the bulge/bar rotates more rapidly than the halo. It is possible that the appearance of rotation in the population of stars that are not confined to the bulge is caused by thick disk stars and bulge stars in the sample which may reach out to distances > 3.5 kpc from the Galactic center. In other words, it is possible that 3.5 kpc is too stringent of a cut and that many stars which participate in the bulge/bar may have apocenters > 3.5 kpc \citep{Portail2017}. However, as noted in Section \ref{sec:confined}, the majority of stars that do not stay confined have eccentricity >0.6 and apocenter > 6 kpc indicating that they are likely halo stars. It is also possible that halo stars that come within 3.5 kpc of the Galactic center have significant prograde rotation. This is not unreasonable given that it has already been observed that halo stars within $\sim$10 kpc of the Galactic center can have prograde rotation up to 50 km/s \citep{Carollo2007}. Furthermore, the confined stars appear to be rotating slower than expectations from the simulation (see right panel of Figure \ref{fig:los}). This has previously been observed among metal-poor bulge stars in \citet{Arentsen2020}. However, since they cannot distinguish between the halo interlopers and confined stars, it is difficult to determine if the slower rotation observed in \citet{Arentsen2020} is a result of halo contamination or the confined bulge stars. Our results indicate that it is in fact the confined stars that rotate slower than expected given the simulations. The slower rotation among confined stars will be discussed further in Section \ref{sec:3dvs}, where we present the rotational velocity ($v_{\phi}$) distribution of confined stars.

In addition, to the differences in rotation, the confined and not confined stars show differences in velocity dispersions. Specifically, the stars that are not confined show much higher velocity dispersions than those which are confined. Our results for all of the stars (left panel of Figure \ref{fig:los}) is consistent with previous work where metal-poor bulge stars have a line-of-sight velocity dispersion of $\sim$100 km/s regardless of Galactic longitude or latitude \citep{Ness2013b,Kunder2016,Arentsen2020}. However, previous studies did not perform orbit analysis and therefore were unable to determine if this high dispersion was indicative of a classical bulge or merely caused by halo interlopers. In the Figure \ref{fig:los}, we show that the velocity dispersion is significantly lower for the confined stars than for the not confined stars. The signature of a B/P bulge is a peak in the velocity dispersions at $l$=0$^{\circ}$ that decreases moving outwards from the Galactic Center. It also generally has lower velocity dispersion moving away from the Galactic plane to higher $|b|$. A classical bulge, on the other hand, would have a velocity dispersion that is independent of Galactic longitude or latitude and would be represented as a horizontal line in Figure \ref{fig:los}. Therefore, our velocity dispersions for the confined stars are consistent with a B/P bulge and there is no need to invoke a classical bulge population.

\subsubsection{3D Velocities}
\label{sec:3dvs}

\begin{figure*}
    \centering
    \includegraphics[width=\linewidth]{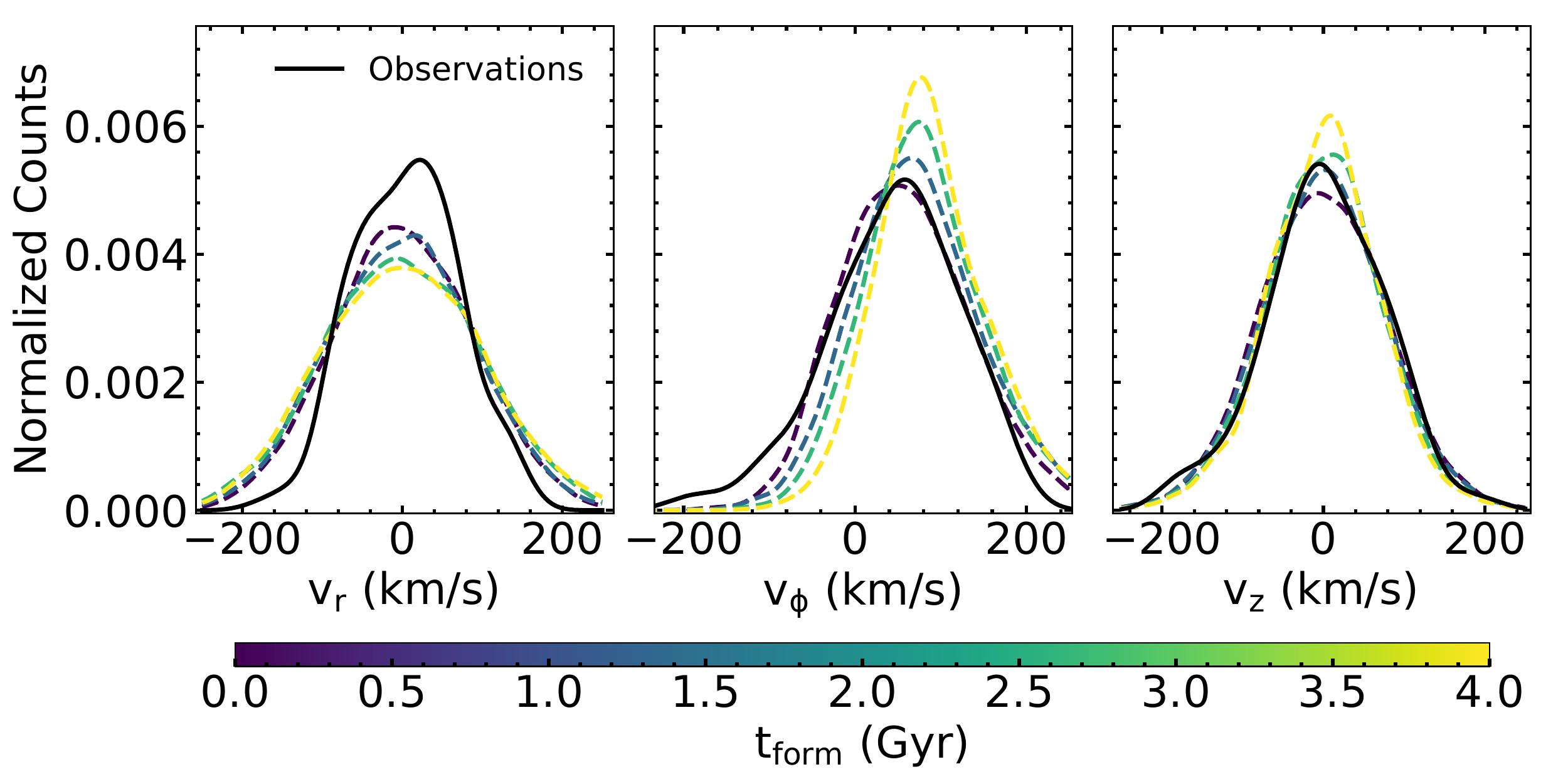}
    \caption{Distribution of the Galactocentric cylindrical velocities for stars with probability of confinement $\geq$50\% (black) compared to populations with different formation times from the simulation (dashed lines) presented in \citet{Cole2014} and \citet{Ness2014}. The distributions from the simulation are determined by using only stars within 3.5 kpc of the Galactic center and along the same line-of-sight as our observations. We only show stars that formed within the first 4 Gyr although the simulation forms stars for all 10 Gyr. Each line is created using 1 Gyr of star formation.}
    \label{fig:allvs}
\end{figure*} 

\begin{figure*}
    \centering
    \includegraphics[width=\linewidth]{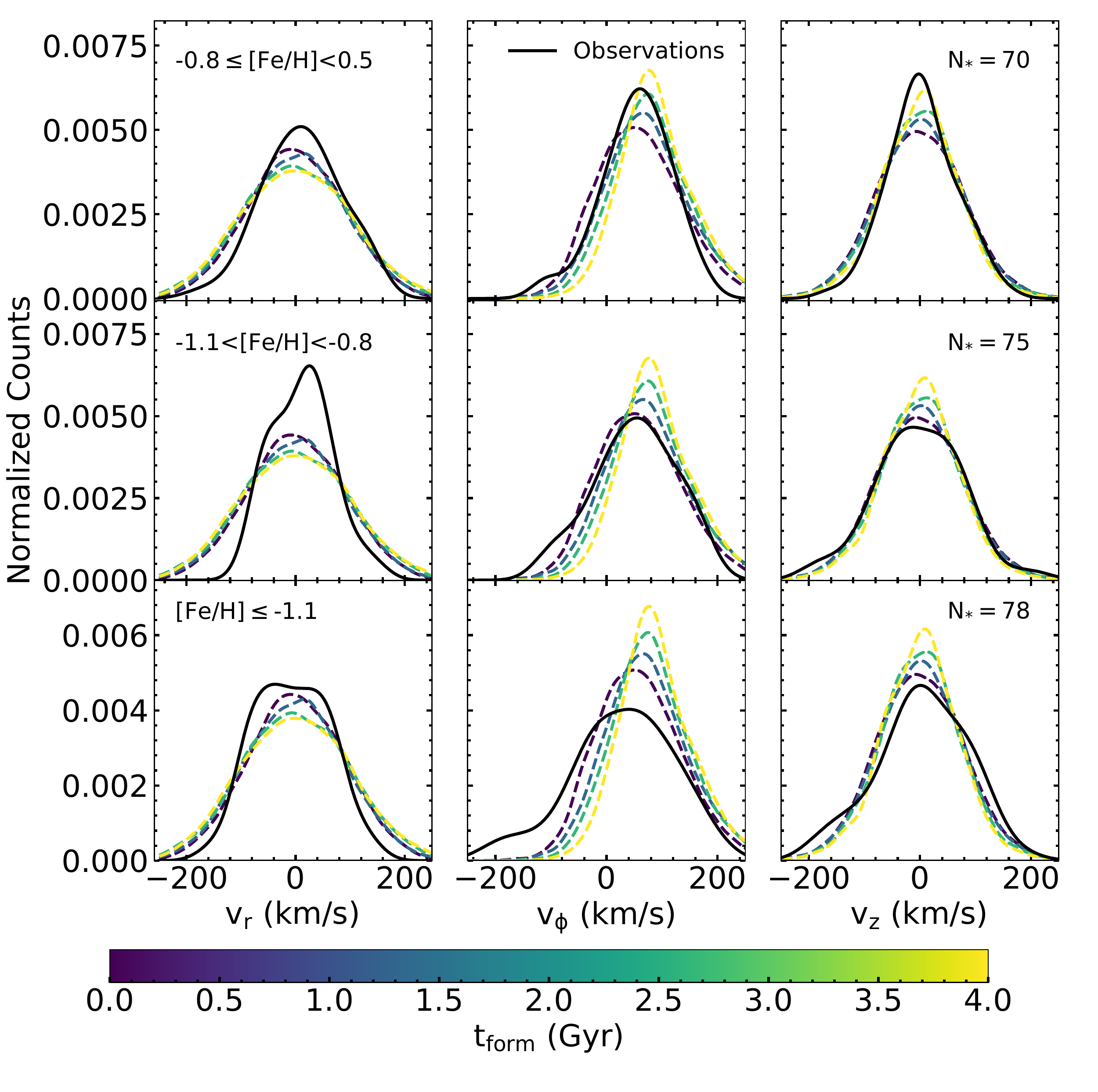}
    \caption{Distribution of the Galactocentric cylindrical velocities for stars with probability of confinement $\geq$ 50\% and varying metallicities (black) compared to populations with different formation times from the simulation (dashed lines) presented in \citet{Cole2014} and \citet{Ness2014}. The simulation lines are the same as those shown in Figure \ref{fig:allvs}. As we move to lower metallicities our observations better match the $v_{\phi}$ distributions for stars that formed earlier with the exception of a growing counter-rotating population. }
    \label{fig:allvs_feh}
\end{figure*}

In addition to the line-of-sight velocities, the full 3D Galactocentric velocities can inform us of the structure and formation history of this population. Specifically, we look at the Galactocentric cylindrical velocities to study the radial motion ($v_r$), rotation ($v_{\phi}$), and vertical motion ($v_z$). In Figure \ref{fig:allvs}, we show the distribution of these velocity components for our stars that have P(conf.) $\geq$ 50\% (black line) along with distributions from the simulation. These distributions are determined by applying a kernel density estimator (KDE) to the observed and simulated distributions. The colored lines shown correspond to populations with different formation times from the simulation. 

The differences in the simulated distributions for different formation ages shown in Figure \ref{fig:allvs} can be explained by kinematic fractionation. Kinematic fractionation, which refers to the separation of populations with different initial kinematics by a growing/forming bar, has been shown to result in older (hence more metal-poor) populations having distinct structure and kinematics that differ from younger (more metal-rich) populations \citep{Debattista2017}. One of the clear trends shown in Figure \ref{fig:allvs} is that the peak of the simulated $v_{\phi}$ distributions approaches zero for stars that formed at earlier times. Therefore, stars that formed earlier generally rotate slower than stars that form later. 

As shown in Figure \ref{fig:allvs}, our observed velocity distributions are mostly consistent with the simulation. However, our observed distribution in $v_r$ is narrower than the simulated distributions. This is likely because we do not confirm that the stars stay confined to within 3.5 kpc of the Galactic center when we calculate the simulated distributions. Therefore, we presumably include more stars with larger $|v_r|$ causing the simulated distributions to be wider than our observed distribution which only includes stars with P(conf.)$\geq$50\%. Additionally, our observed $v_{\phi}$ distribution has a stronger tail of counter-rotating stars than any of the simulated distributions. Specifically, there is a clear overabundance of fast retrograde rotating stars ($v_{\phi}$ <-100 km/s) in our observed distribution. This difference in the distribution likely causes the appearance of slower rotation observed in the right panel of Figure \ref{fig:los} by decreasing the mean line-of-sight velocity. The tails of the $v_z$ distribution are also slightly asymmetrical and differ from the simulation. However, these differences are small and are likely due to stochastic noise in the observed distribution. 

In Figure \ref{fig:allvs_feh}, we show the cylindrical Galactocentric velocity distributions for stars with P(conf.)$\geq$ 50\% divided into three metallicity bins. Additionally, in each panel we show the same simulated distributions as Figure \ref{fig:allvs}. The bins are designed to have similar numbers of stars with the most metal-rich bin having 70 stars, the next bin having 75 stars and the most metal-poor bin having 78 stars. As we move to lower metallicities, the peak of the observed $v_{\phi}$ distribution moves closer to zero. The observed $v_{\phi}$ distribution for the most metal-rich stars (-0.8 dex $\leq$ [Fe/H] < 0.5 dex) is most consistent with the $v_{\phi}$ distribution of stars that formed between 2-3 Gyrs after the start of the simulation. For stars with -1.1 dex $\leq$ [Fe/H] <-0.8 dex, the observed $v_{\phi}$ distribution best matches the simulated distribution for stars that formed between the first 1-2 Gyrs of the simulation. Lastly, the most metal-poor stars ([Fe/H]$\leq$-1.1 dex) have a $v_{\phi}$ distribution similar to the stars that formed within the first Gyr of the simulation. Therefore, our results are consistent with kinematic fractionation if we assume the more metal-poor stars are older. 

We also see a strong counter-rotating tail that is increasingly prominent at lower metallicities that is not in agreement with the simulation distributions. Counter-rotating stars have been observed in the bulge in significant numbers \citep{Queiroz2020}. Although, simulations do predict the presence of some counter-rotating stars in a B/P bulge (see middle panel of Figures \ref{fig:allvs} and \ref{fig:allvs_feh}). Our observations, however, specifically show an overabundance of  stars with $v_{\phi}$<-100 km/s and [Fe/H]$\leq$-1.1 dex, which does warrant further investigation. It is possible that these stars are contamination by halo interlopers, especially given that we have found that the likelihood a star stays confined to the bulge declines with decreasing metallicity (see Section \ref{sec:confined} and \ref{sec:conf_met}). On the other hand, if these stars are bonafide confined bulge stars, it is possible that this is an accreted population. However, they could also be the result of secular evolution, but are not produced in the simulation because of missing physics. For example, the simulation does not include clump formation, which can result in counter-rotating stars \citep{Amarante2020}. In the next installment of this survey we will present the elemental abundances for these stars, which will provide further insights into the origins of these interesting counter-rotating stars. 

Furthermore, the $v_r$ and $v_z$ distributions also  deviate more strongly from the simulation distributions with decreasing metallicity (see Figure \ref{fig:allvs_feh}).  There are a number of factors that may contribute to these deviations. First, as previously discussed, this may be a result of increasing contamination with halo interlopers with decreasing metallicity which is consistent with our results that the frequency of halo interlopers increases with decreasing metallicity (see Figure \ref{fig:frac}). These deviations are also consistent with a possible accreted system that stays within 3.5 kpc of the Galactic center \citep[e.g.,][]{Horta2020}. However, it is also possible that these distributions contain only stars that participate in the B/P bulge and that these deviations are caused by a combination of stochastic noise and varying contributions from different lines-of-sight. Specifically, consistent with a radial and vertical metallicity gradient \citep{Zoccali2008,Gonzalez2011,Johnson2011,Johnson2013a}, the fraction of stars observed at higher $(|l|,|b|)$ becomes larger with decreasing metallicity. On the other hand, the simulated distributions have the highest counts of stars at $(|l|,|b|)$ closer to zero. Therefore, the spatial distribution of the observed sample becomes less similar to the spatial distribution of the simulated sample with decreasing metallicity, which can also cause deviations in velocities, especially in $v_r$. In future work, we will add chemistry information for these stars which will help us distinguish between these possible scenarios.

\section{Discussion and Conclusions}

Many state-of-the-art simulations now indicate that the metal-poor stars in the Galactic bulge are likely to be some of the oldest stars in the Galaxy \citep{Salvadori2010,Tumlinson2010,Starkenburg2017a,El-Badry2018b}. However, in order to determine if these stars are truly ancient, we must understand their origins. For example, it is currently unknown how many, if any, of these stars are confined to the Galactic bulge or are just halo interlopers passing through the bulge. If these stars do stay confined to the bulge, they could participate in the B/P bulge structure or be a classical bulge population. On the other hand, if they are halo interlopers, they could be a unique accreted population \citep[e.g.,][]{Horta2020} or part of the in-situ halo population. The chemodynamical properties of these stars can provide crucial insight into distinguishing between these possible origins. 

Previous work on the metal-poor bulge has mostly been consistent with a classical bulge population. Studies of the chemical make-up of these stars have indicated that they are distinct from halo stars. Specifically, it has been shown that they have lower dispersion and higher Ca abundances than halo stars \citep{Duong2019,Lucey2019} along with differing rates of CEMP stars \citep{Howes2015,Howes2016,Koch2016} and neutron-capture enhanced stars \citep{Koch2019,Lucey2019,Duong2019}. Dynamics of metal-poor bulge stars, specifically the line-of-sight velocities, have indicated that these stars are more consistent with a classical bulge compared to a B/P bulge \citep{Kunder2016}. It has also been shown that the metal-poor stars in the bulge have a higher velocity dispersion than the metal-rich stars, which is inconsistent with a B/P bulge \citep{Ness2013b,Arentsen2020}. These studies also determined that the metal-poor bulge stars rotate slower than the metal-rich stars, which may indicate different origins. However, using N-body simulations, \citet{Gomez2018} demonstrate that a classical bulge population would show even slower rotation than what has been observed among metal-poor bulge stars and that the observations can be explained by a thick disk component. Nevertheless, it is unclear how many, if any of these stars in previous studies are confined bulge stars rather than halo interlopers which are just passing through the bulge. 

There have been a few studies which have performed orbital analysis on metal-poor bulge stars to determine if they stay confined to the bulge. The EMBLA survey found that $\sim$50\% of their sample of very metal-poor stars ([Fe/H]< -2 dex) stay confined to the bulge \citep{Howes2015}. However, only 2 out of the 10 stars that they performed orbital analysis for have apocenters < 3.5 kpc, which we define as the edge of the bulge in this work. Recently, \citet{Reggiani2020} determined that 2 out of the 3 very metal-poor inner bulge stars that they studied have apocenters <3.5 kpc. Finally, only 25\% of the 1389 RR Lyrae stars studied in \citet{Kunder2020} do not stay within 3.5 kpc of the Galactic center. Therefore, the rate at which metal-poor bulge stars stay confined to the bulge varies from 20-75\% depending on selection function.

In this work, we separate the the halo interlopers from the confined metal-poor bulge stars with a probabilistic kinematic method. Using spectra of 523 stars from the VLT/GIRAFFE and VLT/UVES spectrographs along with information \gaia\ DR2 data, we determine the 3D Galactic positions and velocities utilizing a Markov Chain Monte Carlo (MCMC) simulation and Bayesian inference with a Galactic model prior \citep{Rybizki2018}. We then measure the orbital properties and associated errors along with the probability that the star stays confined to the bulge. We also develop a method to derive metallicities from the CaT, which achieves similar precision to previous work \citep{Battaglia2008,Carrera2013} without the need for an estimate of the star's luminosity. We use this method to determine metallicities for the GIRAFFE spectra and also use metallicities determined in \citet{Lucey2019} for the UVES spectra.

Given these data we can conclude: 
\begin{enumerate}
    \item Only $\sim$59\% of the stars in our sample that are currently residing in the bulge have P(conf.)> 50\%.  This value drops to $\sim$14~\% if we only consider stars whose orbits are confined to the bulge with P(conf.) > 90~\%. This indicates that all future and previous studies on the metal-poor bulge that do not perform orbit analysis are likely contaminated by halo stars. 
    \item The rate of contamination with halo interlopers increases with decreasing metallicity. Therefore, it is especially important to perform orbit analysis to separate the halo interlopers from the confined stars when studying stars with [Fe/H] < -2 dex.
    \item By removing the halo interlopers we are able to study the properties of the confined metal-poor bulge stars. We find that the MDF for stars with P(conf.) >90\%  ends at [Fe/H]$\approx$-2 dex. This is consistent with the MDF of the metal-weak thick disk \citep{Beers2014,Carollo2019}.
    \item  We study the kinematics of confined metal-poor bulge stars and find they are consistent with a B/P bulge and kinematic fractionation \citep{Debattista2017}. This is different from previous results, which appeared to be more consistent with a classical bulge because they were unable to remove the halo interlopers  \citep{Ness2013b,Kunder2016,Arentsen2020}.

\end{enumerate}

In the next installment of the COMBS survey, we plan to perform chemical abundance analysis for all 550 GIRAFFE spectra in order to gain further insight on the origins of these stars. For example, we will explore chemical signatures of an accreted population among the stars that do not stay confined to the bulge and test for similarity with the metal-weak thick disk for the stars that do stay confined. We will also search for signatures of globular cluster origins for these stars \citep[e.g.,][]{Schiavon2017}. Combining the dynamical results from this work with chemistry will give us a powerful data set for searching for the oldest stars and studying the origin of the metal-poor bulge population. 

\appendix
\section{Online Table}
We show a section of the available online table in Table \ref{tab:table}. This table includes all 523 stars with 3D positions and velocities in our sample. This table provides the observational properties, estimated metallicities when available, probability of confinement to the bulge, derived distances, 3D Galactic positions and velocities and orbital properties. 

\begin{table*}
\caption{Metallicities, Dynamics, Observational and Orbital properties.}
\label{tab:table}
\begin{tabular}{ccccccccccc}
\hline\hline
Object  & Source ID & l & b & Spec. & SNR & [Fe/H] & RV & $\rm{\sigma_{RV}}$ & P(conf.) & ...  \\
 &  & (deg) & (deg) & & (pixel$^{-1}$) &  & (km/s)& (km/s) & ...\\
\hline
6373.1 & 4085145469159076864 & 14.52 & -9.47 & U & 45 & -1.11$\pm$0.12 & -103.97 & 0.17 & 0.17& ... \\
6382.0 & 4052424582928708096 & 4.47 & -4.81 & U & 64 & -0.82$\pm$0.09 & 106.23 & 0.15 & 0.00 & ...\\
644.0 & 4044703468675705344 & 359.97 & -9.45 & U & 130 & -1.57$\pm$0.06 & 18.68 & 0.9 & 0.91 & ...\\
6531.3 & 4085156464294035968 & 14.60 & -9.62 & U & 48 & -0.98$\pm$0.28 & 192.91 & 0.43 & 0.00 & ...\\
6577.0 & 4085284553070778880 & 15.35 & -9.77 & U & 75 & -0.64$\pm$0.12 & -28.54 & 0.49 & 0.60 & ...\\
6805.0 & 4052435238797214208 & 4.50 & -4.92 & U & 69 & -0.55$\pm$0.37 & -126.34 & 0.68 & 0.68 & ...\\
697.0 & 4044704950509371392 & 0.05 & -9.43 & U & 83 & -1.65$\pm$0.14 & -274.06 & 0.39 & 0.00 & ...\\
7064.3 & 6728167149382096896 & 355.83 & -10.12 & U & 53 & -0.79$\pm$0.15 & -77.31 & 0.61 & 0.54 & ...\\
7362.0 & 4085288577484610048 & 15.29 & -9.58 & G & 29 & ...$\pm$... & -7.94 & 0.45 & 0.00& ... \\
7604.0 & 6728120458827651072 & 355.88 & -10.18 & U & 32 & -1.44$\pm$0.1 & -79.12 & 0.66 & 0.87& ... \\
9094.0 & 4085287718490919424 & 15.44 & -9.74 & U & 73 & -2.31$\pm$0.27 & -54.10 & 0.77 & 0.54 & ...\\
9761.0 & 4085291596814226432 & 15.41 & -9.60 & U & 39 & -0.82$\pm$0.11 & 86.12 & 0.82 & 0.00& ... \\
10036.0 & 6728168493745174528 & 355.89 & -10.09 & G & 45 & -1.36$\pm$0.22 & 11.55 & 2.16 & 0.63 & ...\\
10058.0 & 4085291566781865984 & 15.43 & -9.61 & G & 62 & -0.36$\pm$0.22 & 27.31 & 1.96 & 0.00& ... \\
10073.0 & 6728168317613363200 & 355.88 & -10.05 & G & 28 & -1.18$\pm$0.22 & -58.25 & 2.33 & 0.61& ... \\
10078.0 & 4085291051385691136 & 15.45 & -9.67 & G & 192 & -1.66$\pm$0.22 & -17.43 & 1.20 & 0.12& ... \\
10112.1 & 6728169588922749952 & 355.99 & -10.16 & G & 29 & -0.36$\pm$0.22 & -42.24 & 2.10 & 0.72& ... \\
10123.1 & 6728169039167862784 & 355.93 & -10.15 & G & 56 & -0.97$\pm$0.22 & -0.31 & 0.93 & 0.84& ... \\
10157.1 & 6728128533366200320 & 356.01 & -10.21 & G & 25 & -0.66$\pm$0.22 & -151.08 & 1.89 & 0.49& ... \\
1017.0 & 4044705878222149632 & 0.10 & -9.50 & G & 228 & -0.89$\pm$0.22 & -76.09 & 0.96 & 0.78& ... \\
10205.0 & 4085290948306523648 & 15.45 & -9.64 & G & 108 & -0.96$\pm$0.22 & -33.99 & 0.82 & 0.11 & ...\\
1023.0 & 4044702029931312640 & 0.03 & -9.54 & G & 161 & -0.85$\pm$0.22 & -47.65 & 0.63 & 0.19 & ...\\
10272.0 & 4085290948306547200 & 15.44 & -9.63 & G & 99 & -0.41$\pm$0.22 & 11.78 & 0.82 & 0.00 & ...\\
10301.0 & 4085311461071115264 & 15.51 & -9.77 & G & 112 & -0.37$\pm$0.22 & 102.63 & 1.10 & 0.14 & ...\\
10331.0 & 4085291321939805184 & 15.45 & -9.65 & G & 162 & -0.94$\pm$0.22 & 115.73 & 0.55 & 0.15 & ...\\
...& ...& ...& ...& ...& ...& ...& ...& ...& ... & ...\\
\hline
\end{tabular}
\flushleft{A section of the online table with the given object name, \gaia\ DR2 source ID, Galactic longitude (l) and latitude (b), the spectrograph used to observe (``U" for UVES and ``G" for GIRAFFE), the SNR, estimated [Fe/H] with errors, heliocentric RV with errors, and probability of confinement. Also included in the online table is the distance, Galactic positions (X,Y,Z), Galactic velocities (U,V,W), eccentricity, apocenter, pericenter, and $z_{max}$ along with the associated asymmetric errors for all of these quantities.   }
\end{table*}

\section*{Acknowledgements}
{\small 
We thank Jos de Bruijne and Jan Rybizki for helpful discussions at the International School for Space Science's summer school on ``Space Astrometry for Astrophysics" in 2019.

KH has been partially supported by a TDA/Scialog  (2018-2020) grant funded by the Research Corporation and a TDA/Scialog grant (2019-2021) funded by the Heising-Simons Foundation. KH and ML acknowledges support from the National Science Foundation grant AST-1907417 and from the Wootton Center for Astrophysical Plasma Properties funded under the United States Department of Energy collaborative agreement DE-NA0003843.
V.P.D. is supported by STFC Consolidated grant \#~ST/R000786/1. 
T.B. acknowledges financial support by grant No. 2018-04857 from the Swedish Research Council.
LC is the recipient of an ARC Future Fellowship (project number FT160100402).
C.K. acknowledges funding from the UK Science and Technology Facility Council (STFC) through grant ST/ R000905/1.
AFM acknowledges support from the European Union’s Horizon 2020 research and innovation programme under the Marie Sklodowska-Curie (Grant Agreement No 797100).
The simulation used in this paper was run at the High Performance Computing  Facility of the University of Central Lancashire.

This work has made use of data from the European Space Agency (ESA)
mission {\it Gaia} (\url{https://www.cosmos.esa.int/gaia}), processed by
the {\it Gaia} Data Processing and Analysis Consortium (DPAC,
\url{https://www.cosmos.esa.int/web/gaia/dpac/consortium}). Funding
for the DPAC has been provided by national institutions, in particular
the institutions participating in the {\it Gaia} Multilateral Agreement.}

\section*{Data Availability}
The data underlying this article are available in the ESO Science Archive Facility at \url{http://archive.eso.org/}, and can be accessed with ESO programme ID 089.B-069.

\bibliography{bibliography}
\bsp	
\label{lastpage}
\end{document}